\documentclass[aps,pra,amsmath,notitlepage,nofootinbib,twocolumn,superscriptaddress,10pt]{revtex4-2}
\pdfoutput=1
\usepackage[utf8]{inputenc}
\usepackage[english]{babel}
\usepackage{microtype}
\usepackage[normalem]{ulem}

\usepackage{graphicx}
\usepackage{dcolumn}
\usepackage{bm}
\usepackage{amsmath,amssymb,amsfonts}
\usepackage{booktabs}

\usepackage{braket}
\usepackage{dsfont}
\usepackage{xcolor}
\usepackage{dsfont}
\usepackage{mathtools}

\usepackage{tikz}

\usepackage[breaklinks=true,colorlinks=true,linkcolor=teal,urlcolor=teal,citecolor=teal]{hyperref}
\usepackage{orcidlink}

\newcommand{\rrangle}{\rangle \! \rangle}
\newcommand{\llangle}{\langle \! \langle}

\DeclareMathOperator{\Tr}{Tr}
\let\originalleft\left
\let\originalright\right
\renewcommand{\left}{\mathopen{}\mathclose\bgroup\originalleft}
\renewcommand{\right}{\aftergroup\egroup\originalright}

\begin{document}

\title{A tensor network approach to sensing quantum light-matter interactions}

\author{Aiman Khan}
\thanks{A.~K. and F.~A. contributed equally to this work}
\email{aimankhan1509@protonmail.com}
\affiliation{School of Mathematics and Physics, University of Portsmouth, Portsmouth PO12UP, United Kingdom.}

\author{Francesco Albarelli}
\email{francesco.albarelli@gmail.com}
\affiliation{Scuola Normale Superiore, I-56126 Pisa, Italy.}

\author{Animesh Datta}
\email{animesh.datta@warwick.ac.uk}
\affiliation{Department of Physics, University of Warwick, Coventry CV47AL, United Kingdom.}

\date{\today}

\begin{abstract}

We present the fundamental limits to the precision of estimating parameters of a quantum matter system probed by light, even when some of the light is lost.
This practically inevitable scenario leads to a tripartite quantum system of matter, and light---detected and lost.
Evaluating fundamental information theoretic quantities such as the quantum Fisher information of \emph{only} the detected light was heretofore impossible.
We succeed by expressing the final quantum state of the detected light as a matrix product operator.
We apply our method to resonance fluorescence and pulsed spectroscopy.
For both, we quantify the sub-optimality of continuous homodyning and photo-counting measurements in parameter estimation.
For the latter, we find that single-photon Fock state pulses allow higher precision per photon than pulses of coherent states.
Our method should be valuable in studies of quantum light-matter interactions, quantum light spectroscopy, quantum stochastic thermodynamics, and quantum clocks.
\end{abstract}

\maketitle

\section{Introduction}
\label{sec:introduction}

Quantum light-matter interactions lie at the heart of modern science and technology.
These range from spectroscopy---where the scattered light carries information about the matter system~\cite{mukamelPrinciplesNonlinearOptical1995}, to engineering nonclassical states of light, matter, and combinations thereof.
The latter include novel technological applications such as quantum networks, memories, clocks, as well as novel inorganic~\cite{GonzlezTudela2024} and organic devices~\cite{Khazanov2023}.
Central to studying these various systems are the quantum state of the light before and after the interaction.
The former is typically a coherent state and easily described. 

The scattered light acquires complicated temporal quantum correlations due to the interaction.
Its explicit quantum state is often eschewed in favour of obtaining its emission spectra and correlation functions.
These quantities are typically computed using only the matter (often called emitter) system via tools such as the input-output formalism and the quantum regression theorem~\cite{3540223010,Scully1997,Loudon2000}.
Indeed, all relevant properties of the scattered light can be computed from the matter system alone if no other subsystems are involved, that is, the light-matter system is in a pure bipartite quantum state.
The above tools thus benefit from the Schmidt decomposition for bipartite systems.

A fundamental challenge arises in the presence of a third subsystem, typified, for instance, by the empirical fact that it is impossible to collect all of the scattered light.  
The resulting tripartite pure quantum state of the emitter, the detected, and the lost light subsystems
does not afford a Schmidt decomposition.
Formally, this is because higher-order tensors lack a direct analog to the singular value decomposition for matrices.
This tripartism, coupled with the indefinite number of photons in the incident light, has left the problem of evaluating quantum entropic quantities of the detected light only open until now~\cite{abbasgholinejad2025theory}.

\begin{figure*}
    \includegraphics[width=\textwidth]{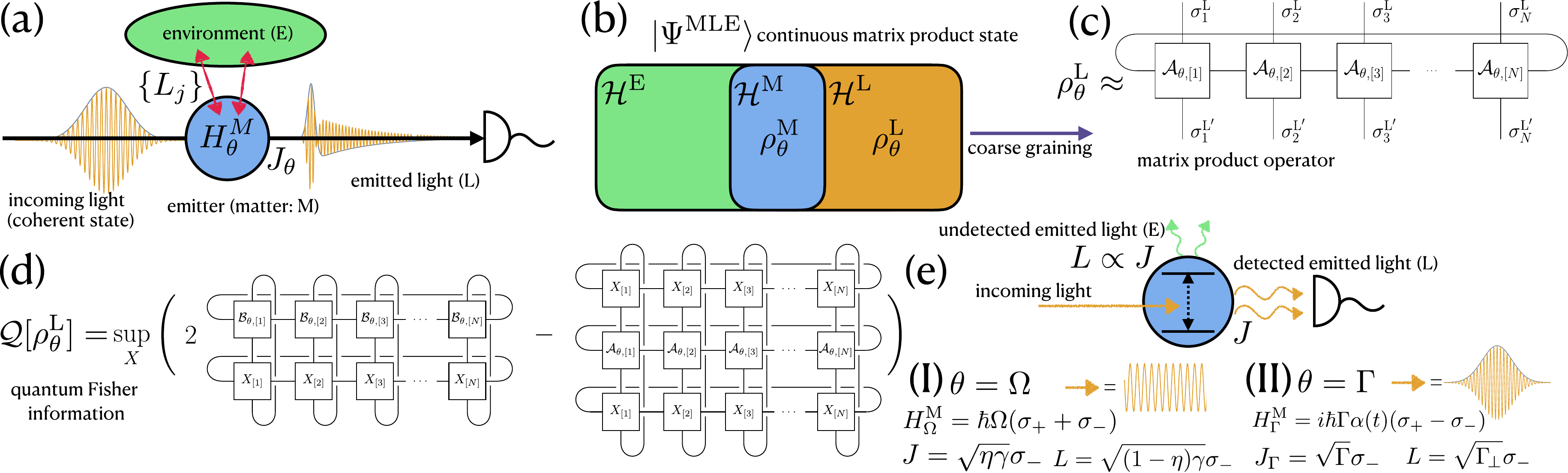}
    \caption{(a) Schematic representation of our tripartite setup. 
    The emitter (M), a matter system with Hamiltonian $H^\mathrm{M}_{\theta}$, interacts with propagating light (L), initially in a coherent state, through an operator $J_\theta$, and also with the environment (E), resulting in the Lindblad operators  $L_j$ acting on M.
    Only L can be detected after the interaction,from which the parameter $\theta$ is to be estimated.
    (b) Tripartism: The tripartite Hilbert space structure $\mathcal{H}^\mathrm{M} \otimes \mathcal{H}^\mathrm{L} \otimes \mathcal{H}^{\mathrm{E}}$.
    The impact of E is purified to an Hamiltonian interaction with operators $L_j$.
    The tripartite state $\ket{\Psi^{\mathrm{MLE}}(t)}$ is a pure continuous matrix product state.
    (c) After coarse-graining into time-bins of size $\Delta t,$ the reduced state of the light $\rho^{\mathrm{L}}_\theta$ is approximately described by a matrix product operator.
    (d) The QFI of a state in MPO form is computed variationally~\cite{Chabuda2020} using Eq.~\eqref{eq:QFI_variational}; the tensors $\mathcal{B}_{\theta,[n]}$ are the MPO representation of the derivative $\partial_\theta \rho^{\mathrm{L}}_\theta$.
    (e) Two case studies considered in this work: (I) Rabi frequency $\Omega$ estimation in resonance fluorescence with detectors of efficiency $\eta <1$; and (II) dipole-moment $\Gamma$ estimation of a two-level system with pulsed classical light, when not all emitted light can be detected. $\Gamma_{\!\perp}$ denotes emission into the lost or undetected modes.}
    \label{fig:scheme}
\end{figure*}

We overcome this by expressing the quantum state of the detected light as a matrix product operator (MPO).
We use it to evaluate the quantum Fisher information (QFI)---a fundamental quantum entropic quantity that sets the precision of estimating matter parameters from the light detected after a quantum light-matter interaction.
To that end, we use a variational optimisation algorithm for states in MPO form~\cite{Chabuda2020}. 
These are our two methodological contributions, schematically depicted in Fig.~\ref{fig:scheme}.
We also evaluate a lower bound to the QFI called the sub-QFI~\cite{Cerezo2021} from the MPO which does not require variational optimisation.
We apply our methods to resonance fluorescence and pulsed spectroscopy with classical light.
The former is a workhorse in fields from quantum optics to quantum clocks~\cite{Singh2025} while the latter is the inevitable benchmark for quantum light spectroscopy~\cite{Dorfman2016,Mukamel2020,Albarelli2023a,Khan2024a, Darsheshdar2024}.

Our main results are as follows:
(i) Obtaining a discrete MPO form for the detected state of the light in Eq.~\eqref{eq:MPDO} and combining it with the variational method in Eq.~\eqref{eq:QFI_variational} to evaluate its QFI;
(ii) Evaluating the QFI for resonance fluorescence with lossy detectors, and quantifying the sub-optimality of continuous homodyne and photodetection measurements therein;
(iii) Showing that, for single-molecule spectroscopies using the same pulse shape, coherent states are less informative than single-photon Fock states on a per-photon basis; 
(iv) Evaluating the sub-QFI from the MPO form, which provides a non-trivial lower bound to the QFI without any optimisation; and
(v) Providing the symmetric logarithmic derivative as an MPO which can inform the construction of measurements attaining the QFI.

While we highlight lossy detection above, the challenge of tripartism emerges whenever the emitter interacts with undetected degrees of freedom.
Going beyond accounting for light emission into undetected modes, our methodological contributions capture general Markovian noise acting on the emitter system.
This includes, for instance, uncorrelated noise of classical origin such as dephasing  due to intensity and phase fluctuations of a driving laser~\cite{Schneider1998}.

Our two methodological contributions of expressing the quantum state of the detected light as an MPO and evaluating a quantum entropic quantity variationally from it should find fruitful applications both individually and in tandem in a broad set of circumstances.
The latter may be used to evaluate other quantum entropic quantities such as entanglement measures for entanglement quantification and manipulation, the quantum relative entropy for quantum hypothesis testing~\cite{Berta2023b} and channel capacities for quantum communication.
Some instances of the former include harnessing the MPO structure to  characterise quantum light sources~\cite{fischerGenerationPulsedQuantum2018},  analysis of quantum networks, the study of driven-dissipative quantum systems~\cite{Marciniak2022} ranging from exciton-polaritons~\cite{Stepanov2019} and quantum clocks to continuously monitored quantum systems~\cite{wiseman2010quantum} and spontaneous symmetry breaking therein~\cite{Adolfo2019}.

By focusing on quantum entropic quantities such as the QFI, our work goes beyond prior uses of tensor network methods in theoretical quantum optics~\cite{Pichler2016,Pichler2017,manzoniSimulatingQuantumLight2017,Droenner2019,Yanagimoto2021,ArranzRegidor2021,Vodenkova2024,Magnifico2025}, as well as the related techniques based on collision models~\cite{Ciccarello2017,Fischer2017,Gross2017,maffeiClosedSystemSolution1D2022}.
Typically, these studied the dynamics of the emitter system, or some specific light observables, or evaluated information theoretic quantities restricted to the bipartite setting noted above.
Some tackled the challenge of tripartism by keeping a pure MPS structure and employing the wavefunction Montecarlo method.
On the contrary, we provide an explicit description of the quantum state of the detected light as a locally purified MPO, a class of states used to study open many-body spin dynamics~\cite{Werner2016a}.
Our work is also independent of tensor-network techniques employed to numerically solve nonperturbative open quantum system dynamics~\cite{Strathearn2018,Somoza2019,Link2024,Fux2024,Cygorek2024b},
and their use in computing optical spectra in the presence of structured phononic environments~\cite{Mascherpa2019,gribben2022exact,Dunnett2021a,deWit2025}.
This was again achieved by computing correlation functions of the emitter without explicit reference to the quantum state of the detected light.

\section{Results}

\subsection{Model description}

\subsubsection{Light-matter interaction}

The matter (M) quantum system, which we call an emitter, is assumed to be pointlike. It interacts via its dipole with a propagating radiation field, described by a one-dimensional continuum of bosonic modes satisfying the commutation relations $[a(\omega),a^\dag(\omega')] = \delta(\omega-\omega')$.
We label the subsystem composed of the bosonic field as the light (L), which will conventionally represent radiation that can be measured.
The Hilbert spaces of the emitter and light are denoted as $\mathcal{H}^\mathrm{M}$ and $\mathcal{H}^\mathrm{L}$.
The free Hamiltonian of the light is
\begin{equation}
    \label{eq:HL}
    H^\mathrm{L} = \int_0^\infty \! d \omega \, \hbar \omega \, a^\dag(\omega) a (\omega),
\end{equation}
and it can be used to model several physical configurations where radiation can be assumed to propagate in one dimension and vectoral properties of electromagnetic field can be ignored, e.g., in a waveguide or in free-space under the paraxial approximation~\cite{Blow1990,deutschParaxialQuantumPropagation1991}.

Under standard Markovian approximations valid in quantum optics~\cite{Scully1997,Loudon2000} (details are provided in Appendix~\ref{app:model} in the Supplemental Material (SM)),
the interaction Hamiltonian in the interaction picture takes the standard form
\begin{equation}\label{eq:Hint_LM}
    H^{\mathrm{ML}}(t) = H^{\mathrm{M}}(t)  +i\hbar(a^{\dag}(t)J  -  a(t) J^\dag ),
\end{equation}
written in terms of the time-domain ``quantum noise'' operators\footnote{$J$ and $a(t)$ have units of the square-root of a frequency and the operator $J$ includes the physical coupling constant.}
\begin{equation}
    \label{eq:a_time}
    a(t) = \frac{1}{\sqrt{2 \pi}} \int \! d\omega \, a(\omega) e^{-i(\omega - \omega_0) t},
\end{equation}
satisfying the commutation relations $[a(t),a^\dag(t')] = \delta(t-t')$.
Here $J$ is the matter-system operator that couples to the field, essentially determined by its dipole moment, and $\omega_0$ is a characteristic frequency frequency of the emitter such that going into the interaction picture, $J$ transforms as $J e^{-i \omega_0 t}$.
The phase factor is thereby absorbed into Eq.~\eqref{eq:a_time} so the operators are centered around $\omega_0$.
Finally, $H^{\mathrm{M}}(t)$ is the interaction picture Hamiltonian of the emitter, which is time-dependent in general and may include a time-dependent classical driving term, we will elaborate on this in Sec.~\ref{subsubsec:Mollow}.

The interaction Hamiltonian~\eqref{eq:Hint_LM} between the emitter and quantum noise operators is tightly connected to the celebrated Gorini–Kossakowski–Sudarshan–Lindblad master equation, simply referred to as master equation (ME) in the following.
As a matter of fact, if the light field is initially in the vacuum, a well-justified assumption at optical frequencies, the reduced state of the emitter obeys the ME
\begin{equation}
    \label{eq:M_Lindblad}
    \frac{d \rho^{\mathrm{M}}(t)}{d t} = -i \hbar \left[ H^{\mathrm{M}}(t) , \rho^{\mathrm{M}}(t) \right] + \mathcal{D}[J]\left(\rho^{\mathrm{M}}(t) \right),
\end{equation}
where we call $\mathcal{D}[\chi](\rho) = \chi \rho \chi^{\dag} - (\chi^{\dag}\chi\rho + \rho\chi^{\dag}\chi)/2$ a Lindblad dissipator\footnote{Despite the nomenclature, the operator $\chi$ may represent other decoherence processes, such as dephasing.}.
This is the reason the light subsystem itself is often labeled as an environment in the literature.
We disown this convention to reserve the term ``environment" (E) for modes of the light field that cannot be detected.
This sets the stage for the methodological advances in this work.

\subsubsection{Markovian noise}
\label{subsec:Markovian_noise}

We say that the dynamics of the emitter is noisy, since it interacts with one or more environments, which, unlike the light field L, are completely inaccessible to experimental observation.
Since the interaction \eqref{eq:Hint_LM} correlates light and matter, the noise acting on the emitter will also influence the reduced state of the light.
In particular, we focus on the case of general Markovian noise, modeled by Lindblad dissipators, so that the overall dynamics of the light-matter system obeys the ME
\begin{equation}
    \label{eq:LM_Lindblad}
    \frac{d \rho^{\mathrm{ML}}(t)}{d t} = -i \hbar \left[ H^\mathrm{ML}(t) , \rho^{\mathrm{ML}}(t) \right] + \sum_{j=1}^{P}\mathcal{D}[L_j](\rho^{\mathrm{ML}}(t)) \, ,
\end{equation}
where the $P < \infty$ operators $L_j$ act only on the matter subsystem.
The validity of adding Lindblad dissipators to the Hamiltonian generator as in Eq.~\eqref{eq:LM_Lindblad} is well-justified in the weak coupling regime~\cite{Brask2017}, our domain of interest.

Regardless of the microscopic origin of the Lindblad dissipators acting on M, it is possible to purify the dynamics by introducing independent bosonic fields $b_j(t)$ satisfying $[b_j(t),b^\dag_k(t')]=\delta_{jk} \delta(t-t')$, one for each $L_j$.
Their collective Hilbert space is denoted as $\mathcal{H}^\mathrm{E}$.
These are initialized in the vacuum.
They interact with the system just like the light field, i.e., a total interaction Hamiltonian
\begin{equation}\label{eq:H_MLE}
    H^{\mathrm{MLE}}(t) = H^{\mathrm{M}}(t)  +i\hbar\left(J a^{\dag}(t) + \sum_{j=1}^P L_j b^{\dag}_j(t) - \mathrm{h.c.}\right).
\end{equation}
In general, this is a useful observation that allows one to treat arbitrary Lindblad operators on the same footing as the coupling operator $J$, even if the fields $b_j(t)$ may not have any real physical meaning.

When describing lost light, the environment is also an optical radiation field and the purified description matches the underlying physics.
This scenario will be relevant for our case studies later, wherein we consider a single ($P=1$) operator $L \propto J$ coupling the emitter to the radiation fields, corresponding to the inevitably  practical scenario in which only a portion of the emitted light corresponding to the modes $a(t)$ can be measured, while the rest of the field in modes $b(t)$ (in which light can be emitted but not detected or measured) acts as the environment E.
This is a manifestation of tripartism, as depicted in Fig.~\ref{fig:scheme} (b).

\subsubsection{Coherent states as classical driving}
\label{subsubsec:Mollow}

We now specify the initial state of the light field L as coherent states, a reasonable assumption at optical frequencies.
For a continuum field, they are defined as $\ket{\alpha^{\mathrm{L}}} = \mathrm{exp}\,\left( \int d\tau \, [\alpha(\tau)a^{\dag}(\tau) - \alpha^{*}(\tau)a(\tau)] \right)\,\ket{0^{\mathrm{L}}}$.
The square-integrable complex amplitude $\alpha(\tau)$ describes a temporal envelope that modulates fast oscillations at the carrier frequency $\omega_0$ (since the state is written in terms of the operators~\eqref{eq:a_time} defined relative to the $\omega_0$).
The integral $\int | \alpha(t)|^2 dt = \alpha^2 = \bar{n}$ represents the total average number of photons, where $\alpha \in \mathbb{R}$ allows us to rewrite the function as $\alpha(\tau) = \alpha \phi (\tau)$ with a normalized complex amplitude $\phi(\tau)$. This will be useful for comparisons with single-photon pulses in Sec.~\ref{subsec:dipole_est}.

Applying a unitary displacement operation transforms the light state into the vacuum while equivalently and simultaneously adding a classical driving term to the system Hamiltonian~\cite[App. C]{Fischer2017}
\begin{equation}
    H^{\mathrm{M}} \mapsto H^{\mathrm{M}} +  i \hbar \left( \alpha^*(t) J - \alpha(t) J^\dag  \right).
\end{equation}
This is known as the Mollow transformation~\cite{mollowPurestateAnalysisResonant1975}.
This displacement acts only on $\mathcal{H}^\mathrm{L}$ and does not depend on any matter or interaction parameters.
Thus, it does not affect quantities such as the QFI of these parameters in the reduced state of the detected light, which are invariant under parameter-independent unitary operations.

The above argument also applies to continuous monochromatic driving, which can be understood as the limit of an infinitely long pulse with an infinite average number of photons, such that the photon flux is fixed.
Therefore, we will always work in the Mollow-transformed picture, having the light subsystem initially in the vacuum and including the classical driving term in the matter system Hamiltonian.

\subsection{State of the light as a matrix product operator}

We capture the evolution of the tripartite MLE system, starting with an initial state $\ket{\Psi^{\mathrm{MLE}}(0)}$ and time-evolving using the interaction Hamiltonian~\eqref{eq:H_MLE}, to $ \ket{\Psi^{\mathrm{MLE}}(t)}.$
This approach is particularly appealing because the formal solution of the Schrödinger equation for $ \ket{\Psi^{\mathrm{MLE}}(t)}$ has the form of a continuous MPS~\cite{Verstraete2010}.
The final state of the light is then 
\begin{equation}
\rho^{\mathrm{L}}(t) = \Tr_{\mathrm{ME}} \left[\ket{\Psi^{\mathrm{MLE}}(t)}\bra{\Psi^{\mathrm{MLE}}(t)}\right].
\end{equation}
Despite this formal solution, evaluating quantum entropic quantities of $\rho^{\mathrm{L}}(t)$ directly in the continuum has remained a challenge.

We overcome this by coarse-graining the time variable, whereby the continuous MPS is approximated by a standard (discrete) MPS.
Our problem of evaluating $ \ket{\Psi^{\mathrm{MLE}}(t)}$ is then mapped to a one-dimensional chain, where the Hilbert spaces of the coarse-grained time bins are the local physical spaces at different sites of the chain.
In this description, the environments are local degrees of freedom at each site of the chain corresponding to each time bin.  
We thus obtain a quantum state of the detected light field $\rho^{\mathrm{L}}(t)$ that is by construction a locally purified MPO~\cite{Verstraete2004a,DelasCuevas2013,Werner2016a}.

\subsubsection{Coarse-graining}
\label{subsec:coarsegraining}

As a first step, the singular correlation structure for the white noise operators $a,b$, defined in Eq.~\eqref{eq:a_time},
requires a coarse-graining, achieved by binning the time interval $[0,t_{\mathrm{fin}})$ in $N$ equally-sized intervals $[n]:= [t_{n-1},t_n),\,\Delta t = t_n - t_{n-1}$, where $t_0 = 0$ and $t_N = t_{\mathrm{fin}} = N\Delta t$.
The corresponding continuous temporal tensor product space of the light is partitioned as $\mathcal{H}^{\mathrm{L}} = \otimes_{n}\,\mathcal{H}^{\mathrm{L}}_n$~(and similarly for the continuous spaces corresponding to each environment E).
The coarse-grained increment operators
\begin{align}
    \Delta A_{[n]} \equiv \int_{t_{n-1}}^{t_n}\,d\tau\,a(\tau),~~
    \Delta B_{j,[n]} \equiv \int_{t_{n-1}}^{t_n}\,d\tau\,b_j(\tau)
\end{align}
on the $n$-th time bin, have the commutation relations
\begin{equation}\label{eq:bincommuation}
    \begin{split}
    [\Delta A_{[n]}, \Delta A_{[n']}^{\dag} ] &= \delta_{nn'}\Delta t, \\ 
    [\Delta B_{j,[n]}, \Delta B_{j',[n']}^{\dag} ] &= \delta_{j,j'}\delta_{nn'}\Delta t.
    \end{split}
\end{equation}
The coarse-grained increment operators $\Delta A_{[n]}$, $\Delta B_{j,[n]}$ (and their corresponding Hermitian conjugates) can be conveniently interpreted as single-mode creation and annihilation operators, so that $\ket{\sigma^{\mathrm{L}}}_n = \frac{(\Delta A^{\dag}_{[n]})^{\sigma}}{\sqrt{\sigma!\Delta t^{\sigma}}}\ket{0^{\mathrm{L}}}_n$ and $\ket{\sigma_j^{\mathrm{E}}}_{n} = \frac{(\Delta B_{j,[n]}^{\dag})^{\sigma}}{\sqrt{\sigma!\Delta t^{\sigma}}}\ket{0_j^{\mathrm{E}}}_{n}$
are discrete $\sigma$-photon Fock states in the $n$-th time bin, where $\ket{0^{\mathrm{L}}}_n$ and $\ket{0^{\mathrm{E}}_j}_n$ are the vacuum kets for the $n$-th bin for the light and environment modes respectively. 

Under this coarse-graining, we expand the exact time evolution generated by the Hamiltonian \eqref{eq:H_MLE} as $U^{\mathrm{MLE}}(t_{\mathrm{fin}}) = T_{\leftarrow} \mathrm{exp}\left[- (i/\hbar) \int_0^{t_{\mathrm{fin}}}\,d\tau H^{\mathrm{MLE}}(\tau)\right]$ to order $O(\Delta t)$ as the  time-ordered, Trotterised product 
\begin{equation}\label{eq:Trotterprod}
    U^{\mathrm{MLE}}(t_{\mathrm{fin}}) = \prod_{n=1}^N\,U^\mathrm{MLE}_{[n]}(\Delta t).
\end{equation} 
Here the time evolution in the time bin $[n]$ is simply $U^\mathrm{MLE}_{[n]}(\Delta t) = \mathrm{exp}(-i\,\int_{t_{n-1}}^{t_n} d\tau\,H^\mathrm{MLE}(\tau)/\hbar). $ The time-ordering has been dropped because of the assumption that $\Delta t$ is small.
It is possible to employ higher-order Trotter expansions, if needed, at this step in order to approximate more faithfully the exact continuous MPO form of the light state as standard, discrete one. However, as we will see, the simple form of the first-order Trotter expansion affords an elegant physical picture, while higher order product formulas are more involved and harder to interpret~\cite{Childs2021}.

\subsubsection{MPO structure with vacuum input}
We consider an initial joint uncorrelated state in $\mathcal{H}^\mathrm{MLE}$ given by
\begin{equation}
    \ket{\Psi^{\mathrm{MLE}}(0)} = \ket{\psi^{\mathrm{M}}(0)}\otimes \ket{0^{\mathrm{L}}} \otimes \ket{0^{\mathrm{E}}},
\end{equation}
where the state of the light is factorized as $\ket{0^{\mathrm{L}}} = \otimes_n \ket{0^{\mathrm{L}}}_n$, where $n=1,\cdots, N$ denotes the time bins.
The same holds for $\ket{0^{\mathrm{E}}}$.
This choice for $\mathcal{H}^\mathrm{L}$ is justified by the assumption that we consider incoming coherent states of the light at optical frequencies, in conjunction with the Mollow transformation (Sec.~\ref{subsubsec:Mollow}) which recast their action as a time-dependent Hamiltonian acting on the emitter.
The initial state of $\mathcal{H}^\mathrm{E}$ is in the vacuum by construction, following the purification argument in Sec.~\ref{subsec:Markovian_noise}.

The outgoing state is obtained by a discrete stroboscopic evolution, in which each entangling unitary $U_{[n]}^{\mathrm{MLE}}$ acts on the emitter and on the $n$-th bin residing in L and E.
This is essentially a standard Markovian collision model~\cite{Ciccarello2022}, in which at each iteration the matter system interacts with a new subsystem, uncorrelated from the preceding subsystems, which in this case is partitioned in L and E, and always initialized in the same state $|0^{\mathrm{L}};0^{\mathrm{E}}\rangle_n$.

The reduced state of the detected light residing in $\mathcal{H}^{\mathrm{L}}$ obtained by tracing out the emitter M and the environment E, then has the form of the following matrix product operator~(MPO), pictorially represented in Fig.~\ref{fig:scheme}~(c)\footnote{In Fig.~\ref{fig:scheme} (c), the initial state of the emitter does not appear explicitly as it amounts to a boundary condition of the MPO which can always be incorporated into the first tensor $\mathcal{A}_{[1]}$.}
\begin{equation}\label{eq:MPDO}
    \rho^{\mathrm{L}}(t_\mathrm{fin}) = \!\!\!  \sum_{ \boldsymbol{\sigma}^{\mathrm{L}}, \boldsymbol{\sigma}^{\mathrm{L}'}}
\!\!\!
\Tr_{\mathrm{M}}\left[ \mathcal{A}_{[N]}^{\sigma_N^{\mathrm{L}},\sigma_N^{\mathrm{L}'}}\dots\mathcal{A}_{[1]}^{\sigma_1^{\mathrm{L}},\sigma_1^{\mathrm{L}'}}\!\!\!\rho^{\mathrm{M}}(0)\right]\!\!
\ket{ \bm{\sigma}^{\mathrm{L}}} \bra{ \bm{\sigma}^{\mathrm{L}'}},
\end{equation}
where $\bm{\sigma}^{\mathrm{L}} \equiv \{\sigma_1^{\mathrm{L}},\cdots,\sigma_N^{\mathrm{L}}\}$ indicates all the physical indices of the MPO, $\rho^{\mathrm{M}}(0) = \ket{\psi^{\mathrm{M}}(0)}\bra{\psi^{\mathrm{M}}(0)}$ is the initial state of the emitter, and $\mathcal{A}_{[n]}^{ \sigma_n^{\mathrm{L}},\sigma_n^{\mathrm{L}'}} \bullet = \sum_{\sigma_n^{\mathrm{E}}} A_{[n]}^{ \{ \sigma_n^{\mathrm{L}},\sigma_n^{\mathrm{E}}\}}\, \bullet \,A_{[n]}^{\{\sigma_n^{\mathrm{L}'},\sigma_n^{\mathrm{E}}\} \dag}$ are superoperators on the $n$-th bin of the emitter M space, resulting in an MPO bond dimension of $D^2$, where $D$ is the Hilbert space dimension of $\mathcal{H}^{\mathrm{M}}$.
In particular, this is a locally purified MPO, evident from the fact that each superoperator $\mathcal{A}_{[n]}^{ \sigma_n^{\mathrm{L}},\sigma_n^{\mathrm{L}'}}$ for the $n$-th time bin contains an independent summation over environment indices $\sigma_n^{\mathrm{E}}$.
The explicit form of the Kraus operators $A_{[n]}^{\{\sigma_n^{\mathrm{L}},\sigma_n^{\mathrm{E}}\}}$ is provided in the Methods, Sec.~\ref{subsec:MPO_details}, together with additional details on this derivation.
While the evolution of the emitter is Markovian, the state of the light \eqref{eq:MPDO} builds up temporal correlations as the evolution progresses.

Notably, the MPO bond dimension depends solely on the dimensionality $D$ of the emitter Hilbert space $\mathcal{H}^{\mathrm{M}}$.
This is a joint consequence of (i) the temporally uncorrelated structure of the incoming coherent state input
\begin{equation}
    \ket{\alpha^{\mathrm{L}}} = \bigotimes_i \ket{\alpha(t_i)},
\end{equation}
where $\ket{\alpha(t_i)} = \mathrm{exp}(\alpha(t_{i-1})\sqrt{\Delta t}\,\Delta A^{\dag}_{[n]} - \mathrm{h.c.})\ket{0^{\mathrm{L}}_i}$ are discrete coherent states on the $n$-th bin of the light state, and 
(ii) the weak light-matter coupling that allows us to express the interaction as Eq.~\eqref{eq:H_MLE}.
Intuitively, the initial state of light carries no temporal correlations across the time-bins (corresponding to an MPO bond dimension of unity), and the weak coupling to the emitter induces only feeble correlations across the time-bins in the outgoing state, corresponding to the relatively small bond dimension of $D^2$.
The small MPO bond dimension of the light state is the basis of the successful application of numerical MPO techniques. 
This is in contrast to, say, squeezed state inputs where the temporal correlations across time-bins~\cite{Gross2022} are expected to enter the form of the outgoing MPO as increased bond dimensions.

Our first key result is the direct MPO form of the detected light in Eq.~\eqref{eq:MPDO}. It will be utilised in subsequent sections to evaluate quantum entropic quantifies heretofore impossible.

\subsection{Precision limits for spectroscopy}
\label{sec:precision limits}

Our main goal in this paper is to investigate the precision limits for the estimation of a physical parameter $\theta$ appearing in the Hamiltonian $H^{\mathrm{ML}}_\theta$, by only measuring the accessible light field L, and not the emitter M and the environments E.
We insert the $\theta$ dependence to emphasise our interest in the estimation of this parameter.
This is the standard approach in quantum spectroscopy~\cite{Albarelli2023a} and also for parameter estimation with continuously monitored quantum systems~\cite{Nurdin2022}.
In this setting, the quantum state that encodes all the accessible information about the physical parameter is thus precisely $\rho^{\mathrm{L}}_\theta(t_\mathrm{fin})$ in Eq.~\eqref{eq:MPDO}
When discussing parameter estimation in subsequent sections, we explicitly highlight the $\theta$ dependence of the state.
In this section, we briefly introduce the relevant quantum entropic quantities involved, deferring more detailed explanations, including definitions, to the Methods.

\subsubsection{QFI of the detected light state}
For a given quantum measurement, mathematically described by a positive operator valued-measure $\{ \Pi_x \geq 0 | \sum_x \Pi_x = \mathbb{I} \}$, the probability of observing the outcome $x$ is $p_\theta(x) = \Tr[ \rho_\theta \Pi_x ]$.
In the limit of many repetitions, the precision in estimating the value of the parameter $\theta$ is captured by the classical Fisher information (CFI) $\mathcal{C}[p_\theta]$, a quantity that depends only on $p_\theta(x)$ and on its derivative $\dot{p_\theta}(x) \equiv d  p_\theta(x) / d \theta = \Tr[\dot{\rho_\theta} \Pi_x ]$, with $\dot{\rho}_\theta \equiv d \rho_\theta / d \theta$.
The CFI is nonnegative and larger values correspond to a better precision in the estimation of $\theta$.
For a given quantum state, the CFI can be maximized over all quantum measurements, obtaining the quantum Fisher information (QFI)~\cite{helstrom1976quantum,Braunstein1994,Paris2009}.
It is a functional of the quantum state $\rho_\theta$ and of its derivative $\dot{\rho}_\theta$.
The QFI can be computed with several equivalent methods, but we use the variational form~\cite{Macieszczak2013}
\begin{equation}
    \label{eq:QFI_variational}
	\mathcal{Q}[ \rho_\theta ] =  \sup_{X = X^\dag} \left\{  2\Tr\left( \dot{\rho}_\theta  X \right)-\Tr\left(\rho_\theta X^2\right)\right\},
\end{equation}
where the optimization is over hermitian operators $X$.
The optimal $X$ is known as the symmetric logarithmic derivative (SLD) operator.

Now the main technical challenge is to compute the QFI of the state $\rho^{\mathrm{L}}_\theta(t_\mathrm{fin})$ in Eq.~\eqref{eq:MPDO}.
To that end, we use Ref.~\cite{Chabuda2020} to translate Eq.~\eqref{eq:QFI_variational} into an efficient variational optimization problem for MPO states.
We begin with an MPO ansatz for $X$, perform an iterative site-by-site, that is, time-bin by time-bin, optimization while increasing the bond dimension of $X$ until convergence is reached.
This is depicted in Fig.~\ref{fig:scheme}(d).
As the derivative $\dot{\rho}^{\mathrm{L}}_\theta(t_\mathrm{fin})$ of an MPO state is not easily written as an MPO with the same (small) bond dimension, we evaluate this numerically, approximated by a finite difference.
More details on this algorithm and its implementation in this work are presented in Appendix~\ref{app:MPO-QFI} in the SM.
We will denote the QFI obtained by this computational method  as ``MPO-QFI'' in the following.

This is our second key result: the evaluation of the MPO-QFI for the quantum state of the detected light in Eq.~\eqref{eq:MPDO} using Eq.~\eqref{eq:QFI_variational}.
We apply it to two physically relevant scenarios of sensing quantum light-matter interactions in Secs.~\ref{subsec:rabi_est}  and~\ref{subsec:dipole_est}.

\subsubsection{Bounds on the QFI of the detected light state}

The calculation of the MPO-QFI outlined above involves a variational optimization, whose computational cost grows with the number of time bins.
It is thus meaningful to seek computationally cheaper lower and upper bound on the QFI that can be computed without any variational optimization.

A lower bound on the QFI, called the sub-QFI $\mathcal{Q}_{\mathrm{sub}}[ \rho^\mathrm{L}_\theta]$ can be obtained~\cite{Cerezo2021} from the MPO using only tensor contractions and no optimizations.
The sub-QFI is based on the super-fidelity~\cite{Miszczak2008}, an upper bound on the fidelity between two quantum states that involves no square roots of the density operators.
It also exploits the fact that the QFI is related to the fidelity between two infinitesimally close states $\rho^{\mathrm{L}}_{\theta}(T)$ and $\rho^{\mathrm{L}}_{\theta+\epsilon}(T)$. 
Crucially, the MPO form in Eq.~\eqref{eq:MPDO} allows a computation of the super-fidelity by tensor contractions alone.
For pure states, the super-fidelity matches the regular fidelity and 
the sub-QFI equals the QFI.
Details on the definition and computation of the sub-QFI can be found in the Methods, Sec.~\ref{subsec:sub-QFI}.

An upper bound on the QFI can be obtained as follows.
When there is no environment E, the global state $\ket{\Psi^{\mathrm{ML}}(t)}$ in Eq.~\eqref{eq:MPSformglobal} is pure, and a simpler method to compute the QFI of this pure state can be used.
This relies on solving a modified version of the ME in Eq.~\eqref{eq:LM_Lindblad}, called a two-sided master equation (TSME)~\cite{Gammelmark2014,Yang2023e}.
Details on the computation of this quantity can be found in the Methods, Sec.~\ref{subsec:TSME}.
Essentially, this is only possible because the system is bipartite as noted in Sec.~\eqref{sec:introduction}.
This method can be applied in the presence of the environment E but this only computes the QFI of the purified state $\rho^{\mathrm{LE}}_\theta = \Tr_{\mathrm{M}} \left[ | \Psi^{\mathrm{MLE}}_\theta \rangle \langle \Psi^{\mathrm{MLE}}_\theta | \right]$, as if the lost light were accessible for measurements.
The QFI of $\rho^{\mathrm{LE}}_\theta$ is a valid upper bound on the QFI of $\rho^{\mathrm{L}}_\theta$ and it will be computed in the following and denoted as ``TSME-QFI''.

\subsubsection{CFI of continuous measurements}

We next consider concrete measurement strategies on the state $\rho^{\mathrm{L}}_{\theta}(t)$ and evaluate their performance by computing their CFI.
Specifically, we focus on two paradigmatic continuous monitoring strategies: photodetection (PD) and homodyne detection (HD).
To understand these measurements, it is easier to work in the coarse-grained picture of Sec.~\ref{subsec:coarsegraining}.
Introducing the proper bosonic operators $ a_{[n]} = \Delta A_{[n]} / \sqrt{\Delta t},$ 
PD corresponds a projection on the eigenstates of $a_{[n]}^\dag a_{[n]}$, that is, a measurement of the occupation numbers of the time bin $[n]$.
In the limit $\Delta t \to 0$, the only possible outcomes for each time bin can be 0 or 1 and the continuous stream of observed outcomes describes a Poisson process.
HD corresponds instead to a projection on the eigenstates of the quadrature operator $e^{-i \varphi} a_{[n]}^\dag + e^{i \varphi} a_{[n]}$, where $\varphi$ decides which quadrature of the field is measured.
In the infinitesimal limit, the observed outcomes correspond to a diffusive stochastic process.
Both these measurements are time-local, since they correspond to projections on states that are separable on the time-bin basis.
These two classes of measurements are not always capable of attaining a CFI that saturates the QFI of $\rho^\mathrm{L}_{\theta}(t)$~\cite{Yang2023e}.

\begin{figure*}
    \includegraphics{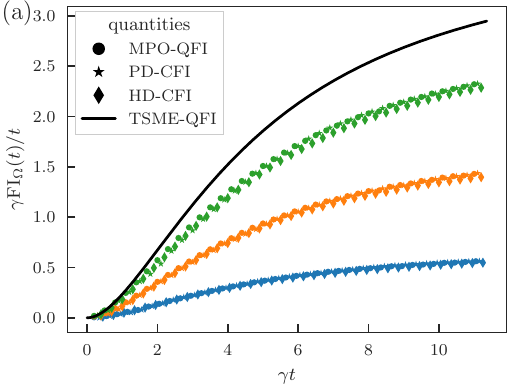}
    \includegraphics{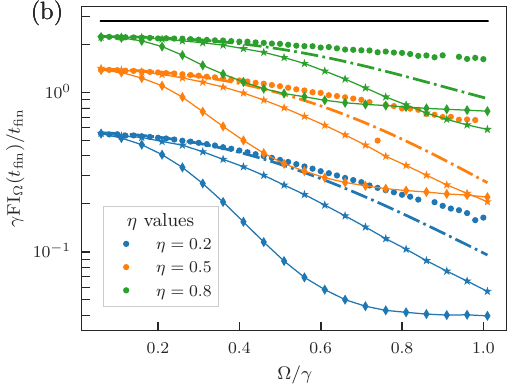}
    \caption{
    QFIs and CFIs for Rabi frequency ($\Omega$) estimation, generically denoted as $\mathrm{FI}_\Omega(t)$ rescaled by the evolution time.
    The markers correspond to different quantities, as in (a); colors correspond to values of the efficiency $\eta$, as in (b).
    Panel (a) shows the dynamics of the various quantities as a function of the evolution time, for a fixed $\Omega = 0.1\gamma$.
    Panel (b) shows the the same figures of merit evaluated at the time $t_\mathrm{fin}=10/\gamma$, as a function of the Rabi frequency; the dash-dotted lines represents the sub-QFI.
    The MPO-QFI is $\mathcal{Q}[ \rho_\Omega^{\mathrm{L}}]$, the TSME-QFI is $\mathcal{Q}[ \rho_\Omega^{\mathrm{LE}} ]$ which for our problem is also the QFI of $\rho^{\mathrm{LE}}_\Omega $ for $\eta=1$, the PD-CFI is $\mathcal{C}[ p_{\Omega}^{\mathrm{pd}} ]$ and the HD-CFI is $\mathcal{C}[p_\Omega^{\mathrm{hd}}]$. The HD-CFI is for a homodyine angle $\varphi = \pi /2$, which is known to be optimal for $\Omega$ estimation~\cite{Kiilerich2016}.
    }
    \label{fig:fluorescence}
\end{figure*}

Although not necessarily optimal, homodyne and photodetection remain typical experimentally.
In both cases, a single realization of the experiment corresponds to a so-called stochastic quantum trajectory of the conditional state of the emitter, which depends on the observed measurement outcomes.
This stochastic dynamics follows Poisson or diffusive stochastic master equations~\cite{wiseman2010quantum,Albarelli2024}, which can be simulated to sample from the PD and HD probabilities $p_{\theta}^{\mathrm{pd}}$ or $p_{\theta}^{\mathrm{hd}}$.
The CFI of these two probabilities, $\mathcal{C}[ p_{\theta}^{\mathrm{pd}} ]$ and $\mathcal{C}[ p_{\theta}^{\mathrm{hd}} ]$ will be dubbed PD-CFI and HD-CFI, respectively.
We compute these quantities using quantum trajectory methods~\cite{Gammelmark2013a,Albarelli2018a,Radaelli2024}.

\subsection{Case study I: Rabi frequency estimation in resonance fluorescence}
\label{subsec:rabi_est}

Resonance fluorescence is a paradigmatic quantum optical phenomenon~\cite{mollowPowerSpectrumLight1969,mollowPurestateAnalysisResonant1975,Kimble1976}, in which a continuous wave laser field in a coherent state illuminates a two-level system (TLS), and the resulting emission field is detected.
The dynamics of the TLS emitter is simple and governed by a ME, also known as Maxwell--Bloch equations for the populations and the coherences. The emission field has a rich spectral-temporal structure, as is evident from the well-known power spectrum~\cite{mollowPowerSpectrumLight1969,Scully1997}. These and other nontrivial aspects continue to be investigated~\cite{LopezCarreno2018,Phillips2020,Hanschke2020,ZubizarretaCasalengua2024,Ngaha2024}.

The excited and ground state of the TLS are denoted as $\ket{e}$ and $\ket{g}$, the lowering and raising operators are $\sigma_{-} = | g \rangle \langle e | $ and $\sigma_{+} = \sigma_{-}^\dag = | e \rangle \langle g | $.
As depicted in Fig.~\ref{fig:scheme}(e), the emitter Hamiltonian (in interaction picture, assuming a monochromatic drive resonant with the TLS), the light-matter coupling operator and the single loss Lindblad operator are 
\begin{equation}
    \begin{split}
   H^\mathrm{M}_{\Omega} &= \hbar\Omega( \sigma_{+} + \sigma_{-} ), \qquad  J = \sqrt{\eta \gamma} \sigma_{-}\\
    L & = \sqrt{ (1-\eta) \gamma } \sigma_{-},
    \end{split}
\end{equation}
where $\gamma \geq 0$ is the overall decay rate and $\eta \in [0,1]$ represents the proportion of light that can be detected, i.e., the measurement efficiency.
We seek to estimate $\theta = \Omega$, the Rabi frequency---a paradigmatic estimation problem ~\cite{Gammelmark2013a,Gammelmark2014,Kiilerich2014,Kiilerich2016,Radaelli2024}.
For no loss ($\eta=1$), PD and HD (with angle $\varphi = \pi/2$) are known to be optimal in the long-time limit.

We present the MPO-QFI for estimating $\Omega$ in Fig.~\ref{fig:fluorescence} for $\eta < 1.$ This has never been evaluated before and is one of our main scientific results.
We also present the CFIs of HD and PD for 
$\eta <1.$
This quantifies the suboptimality of HD and PD, which is due to the intrinsic temporal correlations in the quantum state $\rho_\Omega^\mathrm{L}(t)$.
Our numerical results are for a few representative values of $\eta$. There is only one line corresponding to the upper bound given by the TSME-QFI, since it does not depend on $\eta$.
This also establishes that it is inadequate to study the tripartite problem due to lossy detection.

In panel (a) we plot the quantities $\gamma \mathrm{FI}_\Omega(t) / t$, for a fixed value of $\Omega = 0.1 \gamma$.
The factor $\gamma$ is introduced to provide an adimensional quantity (effectively, this amounts to scaling time and frequency relative to the natural timescale of the problem $\gamma^{-1}$) that illuminates the emergence of a linear behaviour for long times.
This is an expected behaviour of the TSME-QFI when the ME describing the reduced dynamics of the emitter has a unique steady state~\cite{Gammelmark2014}.
Panel (a) also shows that for small Rabi frequencies all the Fisher informations of the state $\rho^\mathrm{L}_\Omega(t)$ (thus excluding the TSME-QFI) are very close to each other. 
This establishes that in this regime continuous photodection or homodyne detection are close to optimal.
In this panel we do not show the sub-QFI, which coincides with the MPO-QFI within numerical error for $\Omega = \gamma/10$.

In panel (b) we fix a final observation time $t_\mathrm{fin} = 10 /\gamma$ and plot the same quantities, i.e. $\gamma \mathrm{FI}_\Omega(t_\mathrm{fin}) / t_\mathrm{fin}$, as a function of the Rabi frequency.
It shows that for larger values of $\Omega$ both HD and PD  are no longer optimal.
This is due to increasing temporal quantum correlations in the state $\rho^{\mathrm{L}}(t_{\mathrm{fin}})$ with larger $\Omega$ that these time-local measurements cannot access. Indeed, their sub-optimality increases with $\Omega.$
Moreover, we show the sub-QFI $\mathcal{Q}_{\mathrm{sub}}[ \rho_\Omega^{\mathrm{L}} ]$ as the dash-dotted line.
It is a tight lower bound for small $\Omega$, but also becomes looser as $\Omega$ increases.
Nevertheless, it is larger than the HD and PD CFIs in Fig.~\ref{fig:fluorescence} (b).

Details on the parameters used in the numerics are provided in Appendix~\ref{app:numericaldetails} in the SM.
Moreover, a figure similar to Fig.~\ref{fig:fluorescence} for two additional efficiency values $\eta = 0.1$ and $\eta=0.99$ is presented in Appendix~\ref{app:additional_results} in the SM. 
We avoid presenting them here for clarity.

\begin{figure*}
    \includegraphics{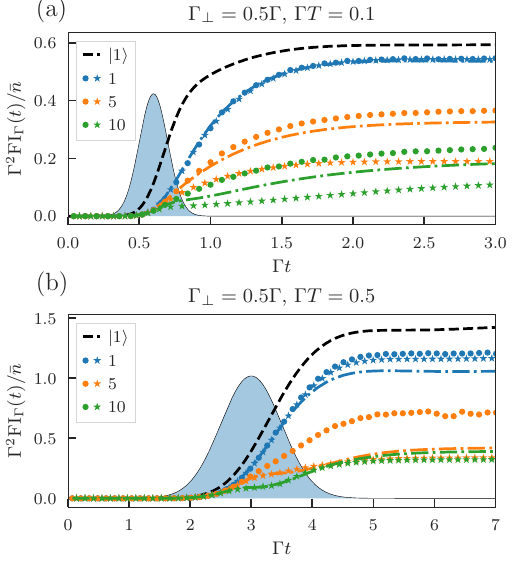}
    \includegraphics{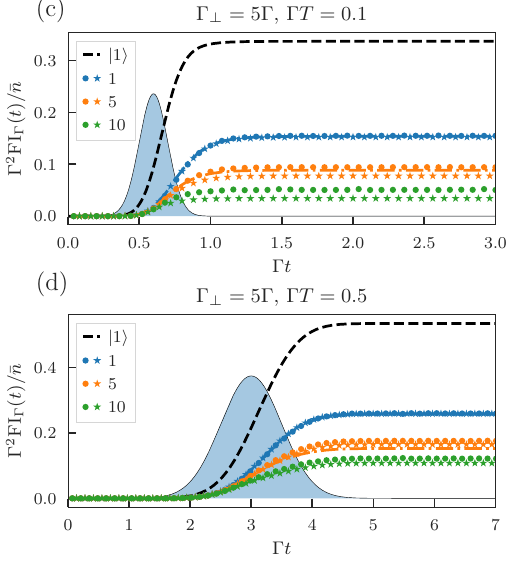}
    \caption{
    QFIs and CFIs for $\Gamma$ estimation with pulsed light, generically denoted as $\mathrm{FI}_\Gamma(t),$ multiplied by $\Gamma^2$ to be adimensional and rescaled by the average number of photons $\bar{n}$ .
   Circles represent the MPO-QFI, stars the PD-CFI, the dash-dotted line the sub-QFI. The colors represent different average number of photons $\bar{n}$ for the coherent states; the black dashed line represents single photons.
    The light-blue shapes is a visual aid that represents the pulse profile $|\phi(t)|^2$ (not to scale on the y-axis). The centre of the pulse is at $t_c = 6.5 T$.
    Each panel corresponds to the parameter values shown on top. 
    To avoid clutter, we only plot the PD-CFI, since HD was found to always perform worse.}
\label{fig:pulsed_Gamma_est}
\end{figure*}

\subsection{Case Study II: Dipole-moment estimation with a coherent state pulse}
\label{subsec:dipole_est}

We now turn to the estimation of the light-matter coupling constant $\Gamma$ between a TLS and a propagating pulse of light in a coherent state.
Since this parameter is proportional to the squared dipole moment of the TLS, we call this problem dipole-moment estimation.
Despite its apparent simplicity, this model has proven to be a valuable testbed for our estimation theoretic approach to quantum spectroscopy, and has been studied in the regime of a single photon interacting with a TLS~\cite{Albarelli2023a,Khan2024a,Darsheshdar2024}.

As depicted in Fig.~\ref{fig:scheme}(e), this setup is related to the previous one, in fact the Rabi frequency $\Omega$ also depends on the TLS dipole moment.
However, it is a different estimation problem, since the parameter to estimate $\theta= \Gamma$ also appears in the emitter-light coupling operator $J.$ Furthermore, 
the pulsed scenario leads to a time-dependent Hamiltonian
\begin{equation}
    \begin{split}
   H^\mathrm{M}_{\Gamma} &= i \hbar \Gamma \alpha(t) ( \sigma_{+} - \sigma_{-} ), \qquad  J_{\Gamma} = \sqrt{\Gamma} \sigma_{-}\\
    L & = \sqrt{\Gamma_{\! \perp}} \sigma_{-}.
    \end{split}
\end{equation}
We again assume the carrier frequency of the pulse to be resonant with the transition frequency of the TLS, and that $\alpha(t) = \alpha \, \phi(t) \in \mathbb{R}$.
This corresponds to the initial coherent state $\mathrm{exp}\left[ \int d\tau \, \alpha(\tau) \left(a^{\dag}(\tau) - a(\tau)\right) \right] \ket{0^{\mathrm{L}}}$.
We focus on a representative Gaussian amplitude
\begin{equation}
    \label{eq:Gaussian_wavefunction}
    \phi(t) = \frac{1}{(2\pi T^2)^{1/4}}\,\,e^{-(t-t_c)^2/(4T^2)},
\end{equation} 
where $T$ is the pulse duration and $t_c$ the arrival time of the pulse.
Finally, we have parametrized the detected and undetected component of the outgoing field using $\Gamma \geq 0 $ and $\Gamma_{\!\perp}$, respectively 
This is a different notation compared to the previous case study but consistent with Refs.~\cite{Albarelli2023a,Khan2024a,Darsheshdar2024}.
The previous parameters are related to the new ones as $\gamma = \Gamma + \Gamma_\perp$ and $\eta = \Gamma/(\Gamma + \Gamma_\perp)$.
The QFI for single photon states was obtained in Ref.~\cite{Albarelli2023a} and its calculation only involves a numerical integration of single variable functions.

We now compare their performance to coherent states in the real-world case of $\Gamma_{\perp} > 0$.
This comparison for $\Gamma_\perp > 0$ would have been impossible without the methodologies we developed.
We find that greater loss in the detection makes the coherent state probe less informative on a per-photon basis.
For a fair comparison, we consider the single photon state $\int d\tau \, \phi(\tau) a^{\dag}(\tau) \ket{0^{\mathrm{L}}}$ with a wavefunction equal to the renormalized amplitude in Eq.~\eqref{eq:Gaussian_wavefunction}.

In Fig.~\ref{fig:pulsed_Gamma_est}, we present the MPO-QFI and the CFI rescaled by the average number of photons $\bar{n} = \alpha^2$ for coherenet states ($\bar{n}=1$ for single photons) for a few combinations of values of $T$ and $\Gamma_\perp$. 
The MPO-QFI shows that on a per-photon basis, single-photon states are more advantageous in estimating $\Gamma$ than coherent states.
This advantage is more marked for larger values of  $\Gamma_\perp$, which matches the intuition from Ref.~\cite{Albarelli2023a} that in this regime the problem becomes similar to a simple single-mode loss estimation, for which Fock states are known to be optimal~\cite{Adesso2009,Nair2018}.
The intuitive explanation from quantum information theory goes as follows: as $\Gamma_\perp$ becomes increasingly larger than $\Gamma$ the light absorbed by the TLS from the coherent state gets emitted almost completely into the environment E.
Thus, for the light the effect of the interaction with the TLS becomes similar to a simple loss of photons from the original single discrete mode $A_\phi = \int d \tau \phi(\tau) a(\tau)$ populated by the coherent state, the perturbation to the modal structure becomes negligible, since the light is emitted mostly into E.

Fig.~\ref{fig:pulsed_Gamma_est} also shows that the QFI per photon falls as the average number of photons of the pulse increases to $\bar{n}=10$.
It also shows that for less energetic pulses, i.e., smaller values of $\bar{n}$, photodetection retrieves most of the QFI.
This ceases to be so as $\bar{n}$ increases.
However, larger values of $\Gamma_\perp$ make the discrepancy between the PD-CFI and the MPO-QFI smaller.
This could also be explained by the intuition that the situation becomes similar to loss estimation as $\Gamma_\perp$ increases, since for that problem photodetection is indeed optimal even for coherent state probes~\cite{Monras2007}.

Finally, Fig.~\ref{fig:pulsed_Gamma_est} shows that, depending on the choice of parameters, the sub-QFI may or may not be a more informative lower bound on the QFI than the PD-CFI.

Details on the parameters used in the numerics are provided in Appendix~\ref{app:numericaldetails} in the SM.
Moreover, an additional figure similar to Fig.~\ref{fig:pulsed_Gamma_est} for a shorter pulse $\Gamma T = 0.01$ is presented in Appendix~\ref{app:additional_results} in the SM.

\section{Discussion}

We have presented an MPO representation of the quantum state of the detected light emitted by a classically driven quantum emitter. This representation in the face of a tripartite light-matter system enabled us to compute its QFI and sub-QFI. 
This had been impossible before.
We believe this to be a novel result, borne out of a fruitful combination of existing techniques.

Our evaluation of the MPO-QFI incurs error from the time discretization $\Delta t$, and to a lesser extent from the approximation of $\dot{\rho}^{\mathrm{L}}_\theta$ with a finite difference.
There is also the possibility of variational optimisation not reaching a universal miniminum.
We found it helpful to rerun the optimisation several times with different initial conditions to avoid this.
However, this increases the computational costs, especially for an increasing $N,$ the number of time bins.
This governed our choices of $\Omega,$ $\bar{n},$ and $t_{\textrm{fin}}$ in Secs. \ref{subsec:rabi_est} and \ref{subsec:dipole_est}.

The space complexity of storing both the quantum state of the detected light in Eq.~\eqref{eq:MPDO} and its parametric derivative scale as $\mathcal{O}(ND^4)$, $D$ being the dimension of $\mathcal{H}^{\mathrm{M}}.$
The evaluation of its QFI requires storing and manipulating the operator $X$ in Eq.~\eqref{eq:QFI_variational}, which has $\mathcal{O}(ND_X^2)$ elements, $D_X$ being the bond dimension of $X$ in a given optimisation loop.

The time complexities of contracting the tensors in Eq.~\eqref{eq:QFI_variational} in each optimisation loop scales as $\mathcal{O}(ND^4D_X^2)$ for the first term of the variation, and $\mathcal{O}(ND^6D_X^2)$ for the second~\cite{catarina2023density}. 
The time complexity of QFI optimisation itself is expected to scale linearly with $N,$ for fixed bond dimension $D_X$ of $X$~\cite{Chabuda2022}.
This suggests a maximum time complexity of 
$\mathcal{O}(N^2D^6D_X^2)$ for evaluating the QFI, where, we have taken the dominant contribution to complexity coming from the second term of the variation in Eq.~\eqref{eq:QFI_variational}.
We empirically find in our numerical optimisations that convergence is always achieved for $D_{X}\leq 8$, for all instances of parameters in Secs. \ref{subsec:rabi_est} and \ref{subsec:dipole_est}.
This leaves the size of the tensor network $N$ as the dominant factor affecting speed of convergence.

The relatively small value of the bond dimension of the optimal $X$ which saturates the variation---the SLD---is indicative of the relatively short-range temporal quantum correlations in the detected light, at least for our choice of parameters.
This suggests that measurements with \textit{limited} time nonlocality spanning a small number of bins may attain the QFI. Such measurements should surpass the performance of homodyne and photodetection.
This is a reflection of the limited amount of temporal entanglement in the detected light state.
This possibility, first raised in \cite{Yang2023e}, can be explored explicitly in our methodology by diagonalising the SLD MPO, whose explicit form we obtain as part of the optimisation.
A fuller discussion of this diagonalisation, achievable using excited-state DMRG~\cite{stoudenmire2012studying}, as well as other exact or numerical methods~\cite{haegeman2017diagonalizing}, is beyond the scope here, and left for future research.

Furthermore, the QFI obtained with our methods can be compared to fundamental bounds in noisy metrology (in presence of E), which are valid when M is accessible and controllable, and can also be coupled to ancillas~\cite{Demkowicz-Dobrzanski2017,Wan2022}.
In this scenario, L can be seen as the ancillas and thus such upper bounds on the QFI apply also to measurements on L only.
It can happen that measurements on L can saturate such fundamental bounds~\cite{Gorecki2024a}, but the conditions under which this happens is, in general, unknown.

An interesting avenue for future research would be to apply our methodology of computing the QFI of the detected light in the presence of a non-trivial phononic environment accompanying the emitter.
This may benefit from tensor network techniques already used for spectroscopy~\cite{Strathearn2018,Dunnett2021a,Link2024,Fux2024,Cygorek2024b}.
Another interesting extension would be to adapt the technique to an input light field that is in a thermal state instead of vacuum.
This would be relevant beyond the optical domain, e.g., for sensors involving microwave fields and superconducting qubits.

The QFI also has applications beyond parameter estimation in quantum stochastic thermodynamics in defining speed limits, thermodynamic and kinetic uncertainty relations~\cite{Hasegawa2019,Hasegawa2021,Nakajima2023,Hasegawa2023a} and quantum clocks~\cite{Singh2025}.
As the TSME-QFI is the only tool employed thus far in such studies, we foresee quantum stochastic thermodynamics benefiting from our methodologies capturing additional Markovian loss.

\emph{Note added:}
During the completion of this work we became aware of a related study by D. Yang \emph{et al.} with a similar goal, but based on a different method.
The submission of both manuscripts to the arXiv has been coordinated.


\section{Methods}
\label{sec:methods}

\subsection{Details on the MPO form of the state}
\label{subsec:MPO_details}
We introduce the Kraus operators
\begin{equation}
    A_{[n]}^{\sigma^{\mathrm{L}}_n,\sigma_n^{\mathrm{E}}} = \prescript{}{n}\langle \sigma^{\mathrm{L}};\sigma^{\mathrm{E}}|U^{\mathrm{MLE}}_{[n]}|0^{\mathrm{L}};0^{\mathrm{E}}\rangle_n,
\end{equation}
where $\ket{\sigma^{\mathrm{E}}}_n = \ket{\sigma_{1}^{\mathrm{E}}}_n\otimes\dots\ket{\sigma_{P}^{\mathrm{E}}}_n$ represents the Fock states on all the $P$ environments corresponding to the different Lindblad operators, while $\sigma_n^{\mathrm{E}} = \{ \sigma^{\mathrm{E}}_{1,n},\dots,\sigma^{\mathrm{E}}_{P,n}\}$ is the corresponding $P$-tuple of occupation numbers.
In terms of the Kraus operators, the explicit outgoing global state is 
\begin{align}\label{eq:MPSformglobal}
    &\ket{\Psi^{\mathrm{MLE}}(T)} = \sum_{\{\sigma_1^{\mathrm{L}},\dots,\sigma_N^{\mathrm{L}}\},{\{\sigma_1^{\mathrm{E}},\dots,\sigma_N^{\mathrm{E}}\}}}  \noindent\\
    &A_{[N]}^{\sigma_N^{\mathrm{L}},\sigma_N^{\mathrm{E}}}\dots A_{[1]}^{\sigma_1^{\mathrm{L}},\sigma_1^{\mathrm{E}}}\,\ket{\psi^{\mathrm{M}}(0)}\otimes\ket{\sigma_N^{\mathrm{L}},\dots,\sigma_1^{\mathrm{L}}}\otimes\ket{\sigma_N^{\mathrm{E}},\dots,\sigma_1^{\mathrm{E}}} \nonumber.
\end{align}
Representing the global state using Kraus operators clearly shows that this is a MPS in the joint basis of the discrete time-bin modes of the light and environment, as well as the matter system state which is propagated in time from left to the right by the action of the time bins.
Its bond dimension is equal to the emitter Hilbert space dimension $D$. 
The state is pictorially represented as ($\sigma^{\mathrm{M}}$ are indices of the emitter space $\mathcal{H}^{\mathrm{M}}$)
\begin{center}
\begin{tikzpicture}[every node/.style={scale=0.6}]
    
    \foreach \x/\c in {0/{1}, 3/{N}}
    {
        \draw (\x -.35,0) -- (\x-.35,1.);
        \draw (\x +.35,0) -- (\x+.35,1.);
        \node at (\x -.35, 1.3) {\LARGE $\sigma_{\c}^{\mathrm{L}}$};
        \node at (\x +.35, 1.3) {\LARGE $\sigma_{\c}^{\mathrm{E}}$};
    }
    
    \draw (-2,0) -- (3,0);
    \draw[line width=1.5mm, white] (1,0) -- (2,0);
    \draw (1.5,0) node {\LARGE $\dots$};
    \draw (3,0) -- (4,0);
    
    \foreach \x/\c in {0/{1}, 3/{N}}
    {
        \draw[fill=white] (\x-0.6,-0.6) rectangle (\x+0.6,0.6);
        \node at (\x,0) {\LARGE $A_{[\c]}$};
    }

    \draw[fill=white] (-2 - 0.6,-0.6) rectangle (-2 + 0.6,0.6);
    \node at (-2,0) {\Large $\ket{\psi^{\mathrm{M}}(0)}$};
    \node at (3+1.35,0.1) {\LARGE $\sigma^{\mathrm{M}}$.};
\end{tikzpicture}
\end{center}
Performing the partial trace of the environment E to obtain the MPO state in Eq.~\eqref{eq:MPDO} corresponds to the following contraction between the MPS in Eq.~\eqref{eq:MPSformglobal} and its hermitian conjugate 
\begin{center}
\begin{tikzpicture}[every node/.style={scale=0.6}]

    \begin{scope}
        \foreach \x/\c in {0/{1}, 3/{N}}
        {
            \draw (\x+.35,0) -- (\x+.35,1) to[out=90, in=90,looseness=2] (\x+0.75,1) to[out=-90, in=90,looseness=2] (\x+0.75,-2.5) to[out=-90, in=-90,looseness=2] (\x+0.35,-2.5) -- (\x+0.35,-1) ;        
            \draw (\x-.35,0.6) -- (\x-.35,1.0);
            \node[anchor=south] at (\x-.35,1.0) {\LARGE $\sigma_{\c}^{\mathrm{L}}$};
        }
    
        \draw (-2,0) -- (3,0);
        \draw[line width=1.5mm, white] (1,0) -- (2,0);
        \draw (1.5,0) node {\LARGE $\dots$};
        \draw (3,0) -- (4,0);
    
        \foreach \x/\c in {0/{1}, 3/{N}}
        {
            \draw[fill=white] (\x-0.6,-0.6) rectangle (\x+0.6,0.6);
            \node at (\x,0) {\LARGE $A_{[\c]}$};
        
        }
    
        \draw[fill=white] (-2-0.6,-0.6) rectangle (-2+0.6,0.6);
        \node at (-2,0) {\Large $\ket{\psi^{\mathrm{M}}(0)}$};
    \end{scope}
    
    \begin{scope}[shift={(0,-1.5)}]  
        \foreach \x/\c in {0/{1}, 3/{N}}
        {
            \draw (\x-.35,-0.6) -- (\x-.35,-1.0);
            \node[anchor=north] at (\x-.35,-1.0) {\LARGE $\sigma_{\c}^{\mathrm{L'}}$};
        }
    
        \draw (-2,0) -- (3,0);
        \draw[line width=1.5mm, white] (1,0) -- (2,0);
        \draw (1.5,0) node {\LARGE $\dots$};
        \draw (3,0) -- (4,0);
    
        \foreach \x/\c in {0/{1}, 3/{N}}
        {
            \draw[fill=white] (\x-0.6,-0.6) rectangle (\x+0.6,0.6);
            \node at (\x,0) {\LARGE $A^*_{[\c]}$};
        }
    
        \draw[fill=white] (-2-0.6,-0.6) rectangle (-2+0.6,0.6);
        \node at (-2,0) {\Large $\bra{\psi^{\mathrm{M}}(0)}$};
    \end{scope}
        \draw (4,0) to[out=0, in=0,looseness=2] (4,-1.5);
\end{tikzpicture}
\end{center}
from which we see that each summation over the environment indices can be applied directly at the corresponding time bin, i.e., the effect of each environment becomes time-local.
From this contraction of indices we obtain the maps $\mathcal{A}_{[n]}^{ \sigma_n^{\mathrm{L}},\sigma_n^{\mathrm{L}'}} \bullet = \sum_{\sigma_n^{\mathrm{E}}} A_{[n]}^{ \{ \sigma_n^{\mathrm{L}},\sigma_n^{\mathrm{E}}\}}\, \bullet \,A_{[n]}^{\{\sigma_n^{\mathrm{L}'},\sigma_n^{\mathrm{E}'}\} \dag}$, introduced in the Results section.

The MPO state of the light in Eq.~\eqref{eq:MPDO} can also be rewritten in Liouville space as
\begin{align}\label{eq:MPDO_vec}
    \rho^{\mathrm{L}}(t_{\mathrm{fin}}) = \sum_{\substack{\sigma_1^{\mathrm{L}},\dots,\sigma_N^{\mathrm{L}} \\ \sigma_1^{\mathrm{L}'},\dots,\sigma_N^{\mathrm{L}'}}} &  |\sigma_N^{\mathrm{L}},\sigma_N^{\mathrm{L}'}\rrangle \otimes\dots,|\sigma_1^{\mathrm{L}},\sigma_1^{\mathrm{L}'}\rrangle   \noindent \\
    &\Tr\left[\mathcal{A}_{[N]}^{\sigma_{N}^{\mathrm{L}},\sigma_{N}^{\mathrm{L}'}}\dots 
    \mathcal{A}_{[1]}^{\sigma_1^{\mathrm{L}},\sigma_1^{\mathrm{L}'}} 
    |\rho^{\mathrm{M}}(0)\rrangle \llangle \mathds{1}|\right] \nonumber
\end{align}
where we have applied a vectorization, mapping the operators $\ket{\sigma_i^\mathrm{L}} \bra{\sigma_j^\mathrm{L}}$ to the vectors $| \sigma_i^\mathrm{L}, \sigma_j^\mathrm{L} \rrangle$, and analogously on the M space, such that the density operator becomes a $D^2$ vector $\rho^{\mathrm{M}}(0) \rightarrow |\rho^{\mathrm{M}}(0)\rrangle$, and the superoperators $\mathcal{A}_{[n}^{\sigma_n^{\mathrm{L}},\sigma_n^{\mathrm{L}'}}$ are now $D^2{\times} D^2$ matrices.
Essentially, Eq.~\eqref{eq:MPDO_vec} is an MPS vectorization of the (mixed) quantum state of the detected light field L.
This state can be pictorially represented as
\begin{center}
\begin{tikzpicture}[every node/.style={scale=0.6}]
    \draw (-1,0) to[out=180,in=180,looseness=2] (-1,0.8) -- (6,0.8) to[out=0,in=0,looseness=2] (6,0);
    
    \foreach \x/\c in {0/{1}, 2/{2}, 5/{N}}
    {
        \draw[line width=1.5pt](\x,0) -- (\x,1.5);
        \node[anchor=west] at (\x+0.1,1.3) {\LARGE $\sigma_{\c}^{\mathrm{L}},\sigma_{\c}^{\mathrm{L}'}$};
    }
    
    \draw[line width=1.5mm, white] (-1,0) -- (5,0);
    \draw (-1,0) -- (3,0);
    \draw (3.5,0) node {\LARGE $\dots$};
    \draw[line width=1.5mm, white] (4,0) -- (6,0);
    \draw (4,0) -- (6,0);

            \foreach \x/\c in {0/{1}, 2/{2}, 5/{N}}
    {
        \draw[fill=white] (\x-0.6,-0.6) rectangle (\x+0.6,0.6);
        \node at (\x,0) {\LARGE $\mathcal{A}_{[\c]}$};
    }
    
\end{tikzpicture}
\end{center}
where the thicker line represents the vectorization,
and the initial state of the matter system M has been absorbed into the site-$1$ tensor for a more concise network representation, such that $\mathcal{A}_{[1]}^{\sigma_1^{\mathrm{L}},\sigma_1^{\mathrm{L}'}} \equiv \mathcal{A}_{[1]}^{\sigma_1^{\mathrm{L}},\sigma_1^{\mathrm{L}'}}|\rho^{\mathrm{M}}(0)\rrangle \llangle \mathds{1}|$.
While we have omitted the explicit parameter dependence here, the superoperators $\mathcal{A}_{[n]}^{\sigma_n^{L} , \sigma_n^{L'} }$ depend on the chosen parameter $\theta$, as they appear in Fig.~\ref{fig:scheme}.

The MPO form of the state $\rho^{\mathrm{L}}(t_{\mathrm{fin}})$ in Eq.~\eqref{eq:MPDO} is a consequence of the collisional structure of the interaction between the emitter M and each time-bin of L and E: at each timestep, M interacts with a new time-bin, always prepared in the same state.
This follows from coarse-graining the unitary evolution as in Eq.~\eqref{eq:Trotterprod} and the assumption that the initial state of L and E is temporally uncorrelated (in particular, the vacuum).
While this structure is general, to proceed concretely, we must specify the form of the Kraus operators describing the light-matter interaction.

Since Eq.~\eqref{eq:Trotterprod} is correct up to order $O(\Delta t)$, for consistency, we seek superoperators $\mathcal{A}_{[n]}^{ \sigma_n^{\mathrm{L}},\sigma_n^{\mathrm{L}'}}$ that are similarly correct up to order $O(\Delta t)$.
To achieve this, it is enough to consider the Kraus operators~\cite{Dalibard1992}
\begin{align}
\label{eq:Kraus_ops}
    &A_{[n]}^{0,\{0,\dots,0\}} = \mathds{1}^{\mathrm{M}} - \frac{i}{\hbar}H^{\mathrm{M}}(t_{n-1})\Delta t - \frac{\Delta t}{2}  (J^{\dag}J + \sum_j L_j^{\dag}L_j), \nonumber\noindent\\
    &A_{[n]}^{1,\{0,\dots,0\}} = J\sqrt{\Delta t},
    \nonumber\noindent\\
    &A_{[n]}^{0,\{0,\dots,1_k,\dots,0\}} = L_k\sqrt{\Delta t}.
\end{align}
Performing this truncation at first order in $\Delta t$ also implies that the dimension of the time-bin Hilbert spaces is truncated to two.
Physically, this is justified for small enough $\Delta t$, such that the probability of having more than one photon in each time-bin can be neglected.
Thus, the indices $\sigma^{\mathrm{L}}_k$ and $\sigma^{\mathrm{E}}_k$ take on values $0$ (no-photon) or $1$ (a photon present in the time-bin) only.

Finally, we also note that if both the light L and environment E subsystems are traced out from the MPS \eqref{eq:MPSformglobal}, and the limit $\Delta t\rightarrow 0$ is taken, we get back the Markovian ME for the emitter dynamics~\cite{Ciccarello2022}
\begin{equation}
\label{eq:ME_LandE}
\frac{\partial}{\partial t}\rho^{\mathrm{M}} = -\frac{i}{\hbar}[H^{\mathrm{M}}(t),\rho^{\mathrm{M}}] + \mathcal{D}[J]\,\,(\rho^{\mathrm{M}}) + \sum_{j=1}^{P}\mathcal{D}[L_j](\rho^{\mathrm{M}}).
\end{equation}
This is unsurprising as the above MPS form obtained from the Hamiltonian dynamics in Eq.~\eqref{eq:H_MLE} is a purification of the model dynamics in Eq.~\eqref{eq:LM_Lindblad}.
However, while the evolution of the emitter is Markovian, the quantum state of the detected light in Eq.~\eqref{eq:MPDO} builds up temporal correlations as the evolution progresses, apparent from the non-zero bond dimension of the vectorised MPS $\rho^{\mathrm{L}}(T)$ in Eq.~\eqref{eq:MPDO_vec}.

\subsection{Basics of quantum estimation theory}
\label{subsec:qet}

Estimation theory studies the precision of estimating the true value of a parameter $\theta$ from experimental observations $x$ sampled from a probability distribution $p_\theta(x)$, belonging to a family characterized by a continuous parameter $\theta$.
In quantum mechanics, the probability distribution is obtained from the Born rule $p_\theta(x)=\Tr\left( \rho_{\theta} \mathrm{\Pi}_{x} \right)$, where the parameter dependence lies only in the state $\rho_{\theta}$ and $\Pi_{x}$ is an element of a positive operator-valued measure (POVM), which mathematically describes a quantum measurement with outcomes labeled by $x$.

The Cramér-Rao bound~\cite{lehmann_theory_1998} states that the variance of an unbiased estimator $\tilde{\theta}(x)$ of $\theta$ satisfies
\begin{equation}
\label{eq:CCRB}
\operatorname{Var}[\tilde{\theta}]
\geq  \frac{1}{\nu\mathcal{C}[p_\theta] },
\end{equation}
where $\nu$ is the number of independent repetitions of the experiment and $\mathcal{C}[p_\theta]$ is the CFI, introduced in Sec.~\ref{sec:precision limits}. It is defined as
\begin{equation} \label{eq:CFI}
	\mathcal{C}[p_\theta] =
	\sum\limits_x p_\theta( x ) \left( \frac{d \log p_\theta(x)}{d \theta}\right)^{2},
\end{equation}
where the summation becomes an integral for continuous distributions.
Unbiased estimators do not always exist, but in the limit $\nu \to \infty,$ the inequality in Eq.~\eqref{eq:CCRB} holds not only for the variance, but also for the mean squared error of any reasonable estimator and it can be saturated~\cite{lehmann_theory_1998}, e.g., by the maximum likelihood estimator.
Thus, the CFI is a classical entropic figure of merit that captures the best precision in the asymptotic limit of many samples.

The QFI employed in Sec.~\ref{sec:precision limits} is obtained by maximizing the CFI over all possible POVMs:~\cite{Braunstein1994,Paris2009}
\begin{equation} 
\label{eq:maxCCRB}
 \mathcal{Q}[\rho_{\theta}] = \max _{\{\Pi_x\}} ~ \mathcal{C}[p_\theta] \, .
\end{equation}
It thus expresses a fundamental quantum limit to the estimation precision, by the following chain of inequalities
\begin{equation} \label{eq:QCRB}
\operatorname{Var}[\tilde{\theta}] \geq \frac{1}{\nu \mathcal{C}(p_\theta) }  \geq \frac{1}{ \nu \mathcal{Q}(\rho_{\theta}) }.
\end{equation}
The latter inequality is known as the quantum CRB.
We always assume that $\nu$ can be made sufficiently large, so that we can meaningfully focus on the CFI and QFI as the relevant figures of merit to quantify the estimation precision.
This setting is known as local estimation, since the CFI and QFI are defined locally around the true value of the parameter ($\rho_\theta$ and $\partial_{\theta} \rho_\theta$ are evaluated at the true value of $\theta$ in all the equations above).

\subsubsection{QFI expressions}

The QFI in Eq.~\eqref{eq:maxCCRB} can also be expressed by a quantum generalization of Eq.~\eqref{eq:CFI}~\cite{helstrom1976quantum,Holevo2011b}
\begin{equation}
    \mathcal{Q}[ \rho_\theta ] = \Tr \left[ \rho_\theta L_{\theta}^2 \right] = \Tr \left[ \frac{ d \rho_\theta }{ d\theta} L_{\theta} \right] \, ,
\end{equation}
where $L_\theta$ is the SLD operator, the optimizer of the maximization in Eq.~\eqref{eq:QFI_variational}, which can be found by solving a Lyapunov equation
\begin{equation}
    \frac{d \rho_\theta }{d \theta} = \frac{ L_\theta \rho_\theta + \rho_\theta L_\theta }{2} \, .
\end{equation}

The QFI is also equivalent to an infinitesimal expansion of the Bures distance~\cite{Braunstein1994,Paris2009}
\begin{equation}
    \label{eq:QFI_fidelity}
     \mathcal{Q}[ \rho_\theta ] = 8 \lim_{\epsilon \to 0} \frac{ 1 - \mathcal{F}[  \rho_\theta , \ \rho_{\theta + \epsilon } ] } { \epsilon^2 } = -4 \frac{\partial^2  \mathcal{F}[  \rho_\theta , \ \rho_{\theta + \epsilon } ] }{\partial \epsilon^2}  ,
\end{equation}
where the fidelity is defined as
\begin{equation}
\label{eq:fidelity}
    \mathcal{F}[ \rho_1 , \rho_2] = \Vert \sqrt{\rho_1} \sqrt{\rho_2} \Vert_1 =\Tr \left[ \sqrt{ \sqrt{\rho_1} \rho_2 \sqrt{\rho_1}  } \right],
\end{equation}
and $\left\Vert O \right\Vert_1 = \Tr\left[ \sqrt{O O^\dag} \right] $ denotes the trace norm of $O$, equal to the sum of the singular values of the matrix $O$.

\subsection{Open system approach to calculate the QFI of light and environment}
\label{subsec:TSME}

As mentioned in the Results section, the overlap between pure tripartite states of the emitter, light and environment can be computed without ever having to explicitly deal with the light and environment Hilbert spaces~\cite{Gammelmark2014,Molmer2015,Macieszczak2016,Cozzini2007}.
This involves solving the dynamics of a modified density operator $\tilde{\rho}_{\theta_1,\theta_2}^{\mathrm{M}}(t)$  satisfying the TSME
\begin{align}
    & \frac{ d \tilde{\rho}_{\theta_1,\theta_2}^{\mathrm{M}}(t)}{d t} =   -i \left( H_{\theta_1}^\mathrm{M}(t)  \tilde{\rho}_{\theta_1,\theta_2}^{\mathrm{M}}(t) -  \tilde{\rho}_{\theta_1,\theta_2}^{\mathrm{M}}(t) H_{\theta_2}^\mathrm{M}(t)    \right) \\ 
    & + J_{\theta_1} \tilde{\rho}_{\theta_1,\theta_2}^{\mathrm{M}}(t) J_{\theta_2}^{\dag} - \frac{1}{2} \left( J_{\theta_1}^\dag J_{\theta_1}  \tilde{\rho}_{\theta_1,\theta_2}^{\mathrm{M}}(t) + \tilde{\rho}_{\theta_1,\theta_2}^{\mathrm{M}}(t) J_{\theta_2}^\dag J_{\theta_2}    \right )  \nonumber \\ 
   & \qquad +  \sum_{j=1}^P \mathcal{D}[L_j] \left( \tilde{\rho}_{\theta_1,\theta_2}^{\mathrm{M}}(t) \right) \nonumber \, .
\end{align}
with the initial condition $\tilde{\rho}_{\theta_1,\theta_2}^{\mathrm{M}}(0) = | \psi^{\mathrm{M}}(0) \rangle \langle \psi^{\mathrm{M}}(0) |.$  The initial matter state $\ket{\psi^{\mathrm{M}}(0)}$ is independent of $\theta$ as it is encoded in the Hamiltonian evolution only.
This equation deviates from the standard ME in Eq.~\eqref{eq:ME_LandE} in that the operators acting on the two sides of $\tilde{\rho}_{\theta_1,\theta_2}^{\mathrm{M}}(t)$ are not the same, but are evaluated for different values of the parameter, justifying the two-sided part of the moniker TSME.
When $\theta_1 = \theta_2 = \theta$ this becomes the standard ME for the reduced state of the emitter $\tilde{\rho}_{\theta}^{\mathrm{M}}(t)$.

From the operator $\tilde{\rho}_{\theta_1,\theta_2}^{\mathrm{M}}(t)$ it is possible to compute the pure-state overlap~\cite{Gammelmark2014,Molmer2015}
\begin{equation}
    \left\langle \Psi_{\theta_1}^{\mathrm{MLE}}(t) | \Psi_{\theta_2}^{\mathrm{MLE}}(t) \right\rangle = \Tr \left[ \tilde{\rho}_{\theta_1,\theta_2}^{\mathrm{M}}(t) \right],
\end{equation}
and also the fidelity between the reduced light-environment states~\cite{Yang2023e}
\begin{equation}
    \mathcal{F}[ \rho^{\mathrm{LE}}_{\theta_1} (t) , \rho^{\mathrm{LE}}_{\theta_2} (t)  ] = \left\Vert \tilde{\rho}_{\theta_1,\theta_2}^{\mathrm{M}}(t) \right\Vert_1 \, .
\end{equation}
This latter result is a direct consequence of the fact that the singular value spectrum of the reduced joint state of L and E is identical to that of the matter system M, following from the Schmidt decomposition.
This enables us to characterise quantum entropic quantities which depend on these singular values (such as the QFI) on the continuous variable space of L and E, using quantities on the smaller M space only.

By the data processing inequality,  tracing out M increases the fidelity $| \left\langle \Psi_{\theta_1}^{\mathrm{MLE}}(t) | \Psi_{\theta_2}^{\mathrm{MLE}}(t) \right\rangle | \leq \mathcal{F}[ \rho^{\mathrm{LE}}_{\theta_1} (t) , \rho^{\mathrm{LE}}_{\theta_2} (t)  ] $ .
Mathematically, this corresponds to the fact that the absolute value of the trace (sum of eigenvalues) is smaller than the trace norm (sum of singular values) of the same operator.
This argument translates to the QFI, by virtue of Eq.~\eqref{eq:QFI_fidelity}, as $\mathcal{Q}\left[ \ket{\Psi_{\theta}^{\mathrm{MLE}}(t)} \right] \geq  \mathcal{Q} \left[  \rho^{\mathrm{LE}}_{\theta}(t) \right]$, consistent with the fact that the QFI is non-increasing under partial tracing.

\subsection{Sub-QFI and super-fidelity of MPO states.}
\label{subsec:sub-QFI}
For two arbitrary states $\rho_1$ and $\rho_2$ the super-fidelity is defined as\footnote{This corresponds to the square root of the original definition in Ref.~\cite{Miszczak2008}; there, also the fidelity is defined as the square of $\mathcal{F}[\rho_1,\rho_2]$ in Eq.~\eqref{eq:fidelity}.}
\begin{equation}
\label{eq:super_fidelity}
    \left( \mathcal{F}_{\mathrm{sup}}[ \rho_1,\rho_2  ] \right)^2 = \Tr \left[\rho_1 \rho_2 \right] + \sqrt{ \left(1- \Tr[ \rho_1^2 ] \right) \left(1- \Tr[ \rho_2^2] \right)},
\end{equation}
which is an upper bound to the Uhlmann fidelity~\cite{Miszczak2008}
\begin{equation}
    \mathcal{F}[ \rho_{\theta_1} , \rho_1 ]  \leq \mathcal{F}_{\mathrm{sup}}[ \rho_1,\rho_2  ] \,.
\end{equation}
The sub-QFI can be defined from the infinitesimal super-fidelity~\cite{Cerezo2021}, analogously to Eq.~\eqref{eq:QFI_fidelity} as
\begin{equation}
    \mathcal{Q}_{\mathrm{sub}}[\rho_\theta] = 8 \lim_{\epsilon \to 0} \frac{ 1 - \mathcal{F}_{\mathrm{sup}}[  \rho_\theta , \ \rho_{\theta + \epsilon } ] } { \epsilon^2 } ,
\end{equation}
which satisfies, by construction
\begin{equation}
    \mathcal{Q}_{\mathrm{sub}}[\rho_\theta] \leq \mathcal{Q}[\rho_\theta] \,.
\end{equation}
The sub-QFI functional is significantly easier to compute than the QFI as the expression in Eq.~\eqref{eq:super_fidelity} does not include any square root of the density operators.
It only depends on the Hilbert-Schmidt product of the two density operators and on their respective purities.
This is particularly valuable for many-body states in MPO form, for which these traces can be obtained as simple tensor contractions, scaling favourably with the length of the MPO network.
On the other hand, the $2$-norm corresponding to Uhlmann fidelities are notoriously hard to compute efficiently for MPO states, whereas they can be computed efficiently for some partitions of the chain~\cite{Hauru2018}.

The Hilbert-Schmidt product between $\rho^{\mathrm{L}}_{\theta}$ and $\rho^{\mathrm{L}}_{\theta + \epsilon}$ is obtained by contracting the tensors as follows
\begin{center}
\begin{tikzpicture}[every node/.style={scale=0.6}]
    \draw (-1,0) to[out=180,in=180,looseness=2] (-1,0.8) -- (6,0.8) to[out=0,in=0,looseness=2] (6,0);
    
    \foreach \x/\c in {0/{1}, 2/{2}, 5/{N}}
    {
        \draw[line width=1.5pt](\x,0) -- (\x,1.5);
        \node[anchor=west] at (\x+0.1,1.1) {\LARGE $\sigma_{\c}^{\mathrm{L}},\sigma_{\c}^{\mathrm{L}'}$};
    }
    
    \draw[line width=1.5mm, white] (-1,0) -- (5,0);
    \draw (-1,0) -- (3,0);
    \draw (3.5,0) node {$\dots$};
    \draw[line width=1.5mm, white] (4,0) -- (6,0);
    \draw (4,0) -- (6,0);
    
        \foreach \x/\c in {0/{1}, 2/{2}, 5/{N}}
    {
        \draw[fill=white] (\x-0.6,-0.6) rectangle (\x+0.6,0.6);
        \node at (\x,0) {\LARGE $\mathcal{A}_{\theta,[\c]}$};
    }

    \draw (-1,2) to[out=180,in=180,looseness=2] (-1,2+0.8) -- (6,2+0.8) to[out=0,in=0,looseness=2] (6,2);

    \draw[line width=1.5mm, white] (-1,2) -- (5,2);
    \draw (-1,2) -- (3,2);
    \draw (3.5,2) node {$\dots$};
    \draw[line width=1.5mm, white] (4,2) -- (6,2);
    \draw (4,2) -- (6,2);

        \foreach \x/\c in {0/{1}, 2/{2}, 5/{N}}
    {
        \draw[fill=white] (\x-0.6,2 -0.6) rectangle (\x+0.6,2 + 0.6);
        \node at (\x,2) {\LARGE $\mathcal{A}^*_{\theta+\epsilon,[\c]}$};
    }
\end{tikzpicture}
\end{center}
and analogously to compute the purity of an MPO, allowing us then to evalute the sub-QFI using tensor contractions only.

\subsection{Classical Fisher information for continuous photodetection and homodyne detection}
\label{subsec:CFItraj}

When a time-local measurement of the light is performed, it is possible to describe the evolution of the conditional state of the emitter, which depends on the observed measurement outcomes, by means of a stochastic master equation (SME), i.e., a stochastic differential equation for the conditional density matrix of state.
This takes into account the measurement back-action on the state of the emitter and the stochastic nature of the measurement outcomes.
The two most paradigmatic and physically relevant time-local measurements are continuous photodetection and homodyne detection~\cite{wiseman2010quantum,Albarelli2024}, which lead to Poisson and diffusive stochastic differential equations respectively.

For clarity, we use $\varrho^\mathrm{M}$ instead of $\rho^\mathrm{M}$ for the conditional state of the emitter, which depends on the past outcomes of the continuous light measurement.
We also omit the $\theta$-dependence for brevity.
The SME for photodetection then reads
\begin{align}
& d\varrho^{\mathrm{M}} = - i [H^{\mathrm{M}}(t)  , \varrho^{\mathrm{M}}]\,dt + \sum_{j=1}^P \mathcal{D}[L_j] \varrho^{\mathrm{M}} dt \notag \\
& -\frac{1}{2} (J^\dag J \varrho^{\mathrm{M}} + \varrho^{\mathrm{M}} J^\dag J ) \,dt +  \Tr[\varrho^{\mathrm{M}} J^\dag J ]\varrho^{\mathrm{M}} \,dt \notag \\
& + \left( \frac{ J \varrho^{\mathrm{M}} J^\dag}{\Tr[\varrho^{\mathrm{M}} J^\dag J ] } - \varrho^{\mathrm{M}} \right)dN   \,, \label{eq:photoSME} 
\end{align}
where $dN$ is a Poisson increment with state-dependent mean:
\begin{equation}
    \mathbb{E}[dN ] = \Tr[J^\dag J \varrho^{\mathrm{M}}]\,dt \,.
\end{equation}
The SME for homodyne detection reads
\begin{align}
d\varrho^{\mathrm{M}} ={}& - i [H^{\mathrm{M}}(t) , \varrho^{\mathrm{M}}]\,dt + \sum_{j=1}^P \mathcal{D}[L_j] \varrho^{\mathrm{M}} dt  \notag \\
& + \mathcal{D}[J] \varrho^{\mathrm{M}} \,dt + \mathcal{H}[J e^{i \varphi} ] \varrho^{\mathrm{M}} \, dw \:,\label{eq:homoSME}
\end{align}
where  $\mathcal{H}[ J ] \bullet = J \bullet + \bullet J^\dag - \Tr[ \bullet (e^{i \varphi} J +  e^{-i \varphi} J^\dag) ] \bullet $, and
 $dw = dy - \Tr[\varrho^{\mathrm{M}} ( e^{i \varphi} J +  e^{-i \varphi} J^\dag )]$ represents a standard Wiener increment (s.t.
$ (dw)^2 = dt$), operationally corresponding to the difference between the measurement result $dy$ and the rescaled average value of the operator $(J e^{i \varphi} + e^{-i \varphi} J^\dag)$.
Lossy detection corresponds to a rescaling of the light operator $ J \mapsto \sqrt{\eta} J $ and to the introduction of an additional noise operator $ L = \sqrt{1-\eta} J $.

Notice that these SMEs are nonlinear, due to the need of renormalizing the quantum state during the evolution.
Alternatively, one can also consider the evolution of the unnormalized conditional state $\tilde{\varrho}^{M}$ which satisfies a linear SME~\cite{Gammelmark2013a,wiseman2010quantum}.
The trace of this stochastically evolving unnormalized state corresponds (modulo a proportionality constant, irrelevant for parameter estimation) to the probability of observing the corresponding measurement outcomes.
Thus, from this probability one could compute the corresponding CFI.
However, this is impractical for concrete numerical calculations, because the trace of $\tilde{\varrho}^{M}$ becomes very small.
A more numerically stable approach is to consider the so-called monitoring operator
\begin{align}
\tau= \frac{ \partial_\theta \tilde{\varrho}^{M} }{\Tr[\tilde{\varrho}^{M}]} \:, \label{eq:tau_def}
\end{align}
which satisfies a second stochastic equation, coupled to the SME for $\varrho^{\mathrm{M}}$, and suitable for numerical solution.
The CFI can be computed by averaging $\Tr[ \tau ]$ over many stochastic realizations of the SME, i.e., by simulating many possible series of measurement outcomes.
Concrectely, we apply this formalism~\cite{Gammelmark2014} in the refined form of Ref.~\cite{Albarelli2018a}, where it was combined with the idea of simulating the SME with a method that preserves the positivity of the conditional state~\cite{Rouchon2015}.

\vspace{0.3cm}
\begin{acknowledgments}
    We thank Elnaz Darsheshdar and Sourav Das for many discussions on quantum light spectroscopy.
    A.~D. thanks Jonathan Gross, Gerard Milburn, and Kavan Modi for interesting discussions.
    F.~A. thanks Dayou Yang for useful discussions.
    F.~A. acknowledges financial support from Marie Skłodowska-Curie Action EUHORIZON-MSCA-2021PF-01 (project QECANM, grant n. 101068347).
    This work has been funded, in part, by an EPSRC New Horizons grant (EP/V04818X/1) and the UKRI (Reference Number: 10038209) under the UK Government’s Horizon Europe Guarantee for the Research and Innovation Programme under agreement 101070700 (MIRAQLS).
    Computing facilities were 
    provided by the Scientific Computing Research Technology Platform of the University of Warwick.
\end{acknowledgments}

\bibliography{biblio}

\clearpage
\appendix
\onecolumngrid

\begin{center}
\begin{Large}
    Supplemental Material
\end{Large}
\end{center}

\section{Microscopic derivation of the light-matter interaction model}
\label{app:model}

The dynamics of a noisy light-matter system can be described in the Hamiltonian framework, partitioned as
\begin{equation}\label{eq:Hamiltonianfram}
    H = H^{\mathrm{M}} + H^{\mathrm{L}} + H^{\mathrm{E}} + H^{\mathrm{MLE}},
\end{equation}
where $H^{\mathrm{M}}$ is the Hamiltonian characterising the internal dynamics of the quantum emitter under study,
\begin{align}
    H^{\mathrm{L}} &= \int_0^{\infty} d\omega\, \hbar\omega a^{\dag}(\omega)a(\omega);~~ H^{\mathrm{E}} = \sum_{j=1}^{P} \int_0^{\infty} d\omega_j \hbar\omega_j b_j^{\dag}(\omega_j)b_j(\omega_j) 
\end{align}
are $P+1$ (bosonic) free-field Hamiltonians corresponding to detected (denoted as $H^{\mathrm{L}}$) and undetected (denoted as $H^{\mathrm{E}}$) environmental modes, and 
\begin{align}
    H^{\mathrm{MLE}} = &-i\hbar\int_{0}^{\infty} d\omega \,\,\kappa(\omega)[J+J^{\dag}][a(\omega)-a^{\dag}(\omega)] 
    -i\hbar\sum_{j=1}^{P} \int_{0}^{\infty} d\omega_j\,\, \eta_j(\omega_j)[L_j+L_j^{\dag}][b_j(\omega_j)-b^{\dag}_j(\omega_j)]
\end{align}
are the interaction couplings between M and the bosonic environments, with $\{\kappa(\omega),\eta_j(\omega_j)\}$ quantifying the frequency response of the emitter M to the respective modes.
In practical implementations, the measured channel is typically of electromagnetic nature, the simplest and most pragmatic example being laser-driven quantum emitters~\cite{wiseman2010quantum}.
In contrast, the unmonitored bosonic modes may be electromagnetic (such as undetected portions of outgoing light) or of phononic origin. 

We will confine our discussion here to bosonic environmental modes that induce unconditional emitter dynamics that are Lindbladian, noting only  that a more complicated, and possibly non-Markovian, dynamics induced by the most general environmental modes may be dealt with a more involved application of our formalism.
In order to understand more transparently what the Markovian assumption entails, we first employ the rotating wave approximation in the interaction picture generated by the zeroth Hamiltonian,
\begin{equation}
    H_0 = H^{\mathrm{M}}_0 + H^{\mathrm{L}} + H^{\mathrm{E}},
\end{equation}
where we ensure that $H^{\mathrm{M}}_0$ is independent of any parameter to be inferred from the measurement record, in order for the fundamental limits of parameter estimation that we seek to perform on the monitored system to be frame-independent. The coupling Hamiltonian then has the form
\begin{equation}\label{eq:HamiltonianframT}
    H(t) = H^{\mathrm{M}}(t)  +i\hbar(Ja^{\dag}(t) + \sum_{j=1}^P L_j b^{\dag}_j(t) - \mathrm{h.c.})
\end{equation}
where $H^{\mathrm{M}}(t) = \mathrm{exp}(iH_0^{\mathrm{M}}t/\hbar)\,H^{\mathrm{M}}\,\mathrm{exp}(-iH_0^{\mathrm{M}}t/\hbar)$ is the emitter Hamiltonian in the interaction picture,
\begin{equation}
    [a(t),a^{\dag}(t')] = \int_0^{\infty} d\omega \left(\frac{\kappa(\omega)}{\kappa(\omega_0)} \right)^2 \,\frac{e^{-i(\omega-\omega_0)(t-t')}}{2\pi},
\end{equation}
$\omega_0$ is a characteristic emitter frequency, and
\begin{equation}\label{eq:halfFourier}
    a(t) = \frac{1}{\sqrt{2\pi}\kappa(\omega_0)}\,\,\int_0^{\infty} d\omega \kappa(\omega) a(\omega) e^{-i(\omega-\omega_0)t}.
\end{equation}
Similar commutation relations will hold for unmonitored operators $\{b_j(t)\}$, belying the nature of emitter-mode coupling for each respective channel.

In the next, Markovian step, the emitter M is assumed to have a flat spectral response to \emph{all} environmental modes, meaning that one can make the substitution $\kappa(\omega) \rightarrow \kappa(\omega_0)$, and extend the lower limits of integration in Eq. \ref{eq:halfFourier} to $-\infty$, yielding 
\begin{equation}\label{eq:whitenoiseop}
    [a(t),a(t')] = \delta(t-t'), ~~[b_j(t),b_k^{\dag}(t')] = \delta_{jk}\,\delta(t-t').
\end{equation}
The above white-noise correlations of what are now the Fourier transforms of the field operators $a(\omega),\{b_j(\omega)\}$ are intricately tied to the eventual Lindbladian dynamics they induce on the quantum emitter.
In fact, the white-noise operator correlations for bosonic baths indicated by Eq.~\eqref{eq:whitenoiseop} are a necessary and sufficient condition for the unconditional dynamics of the emitter system to be Lindblad.
In retrospect, we can exploit this equivalence to argue that the treatment of unconditional quantum dynamics of an emitter with a single monitored and $N$ unmonitored channels, as given in Eq.~\eqref{eq:LM_Lindblad}, can be adequately described in the most general sense in the Hamiltonian framework of Eq.~\eqref{eq:HamiltonianframT}.

\section{Variational algorithm for QFI calculation of MPO states}
\label{app:MPO-QFI}

To evaluate the QFI using the MPO form of the detetced light state in Eq.~\eqref{eq:MPDO}, we use a variational algorithm~\cite{Chabuda2020,Chabuda2022}, which we describe briefly for completeness.

The algorithm in Ref.~\cite{Chabuda2020} is rooted in the variational form of the QFI in Eq.~\eqref{eq:QFI_variational}, an optimisation over an Hermitian operator $X$.
For the many-body state of light in Eq.~\eqref{eq:MPDO}, we employ the following MPO ansatz for the Hermitian operator $X$ that must be optimised,
\begin{equation}
\label{eq:sldMPO}
    X = \!\!\!  \sum_{ \boldsymbol{\sigma}^{\mathrm{L}}, \boldsymbol{\sigma}^{\mathrm{L}'}}
\!\!\!
\Tr \left[ X_{[N]}^{\sigma_N^{\mathrm{L}},\sigma_N^{\mathrm{L}'}}\dots X_{[1]}^{\sigma_1^{\mathrm{L}},\sigma_1^{\mathrm{L}'}}\,\,\,\!\!\!\right]\!\!
\ket{ \bm{\sigma}^{\mathrm{L}}} \bra{ \bm{\sigma}^{\mathrm{L}'}},
\end{equation}
where the bond-dimension of to the SLD-MPO, corresponding to the internal matrix dimensions of the set of matrices $\{X_{[1]},\dots,X_{[N]}\}$, is fixed at some value $D_{\mathrm{X}}$. We also require
\begin{equation}
X_{[n]}^{\sigma_n^{\mathrm{L}},\sigma_n^{\mathrm{L}'}} = \left(X_{[n]}^{\sigma_n^{\mathrm{L}'},\sigma_n^{\mathrm{L}}}\right)^* 
\end{equation}
to ensure hermiticity.
This does not specify the matrix elements uniquely, but rather a gauge choice for the MPO, referred to as the Hermitian gauge. 
Next, we use the finite difference formula for matrix derivative 
\begin{equation}\label{eq:finitdif}
     \frac{\partial \rho_\theta}{\partial \theta} = \frac{\rho(\theta+\delta) - \rho(\theta-\delta)}{2\delta}
\end{equation}
to construct the MPO form for the derivative of the light state
\begin{equation}
\label{eq:drhoMPO}
   \frac{\partial \rho^{\mathrm{L}}_\theta(t_\mathrm{fin})}{\partial \theta}   
   = \!\!\!  \sum_{ \boldsymbol{\sigma}^{\mathrm{L}}, \boldsymbol{\sigma}^{\mathrm{L}'}}
\!\!\!
\Tr \left[ \mathcal{B}_{[N]}^{\sigma_N^{\mathrm{L}},\sigma_N^{\mathrm{L}'}}\dots \mathcal{B}_{[1]}^{\sigma_1^{\mathrm{L}},\sigma_1^{\mathrm{L}'}}\,\,\,\!\!\!\right]\!\!
\ket{ \bm{\sigma}^{\mathrm{L}}} \bra{ \bm{\sigma}^{\mathrm{L}'}},
\end{equation}
where the tensor elements of the derivative MPO are given as the following direct sum~\cite{Schollwock2011}
\begin{align}   
\label{eq:derMPO}
\mathcal{B}_{[n]}^{\sigma_n^{\mathrm{L}},\sigma_n^{\mathrm{L}'}} &= \frac{1}{(2\delta)^{1/N}}~\mathcal{A}_{[n]}^{\sigma_n^{\mathrm{L}},\sigma_n^{\mathrm{L}'}}(\theta+\delta) \oplus(-\mathcal{A}_{[n]}^{\sigma_n^{\mathrm{L}},\sigma_n^{\mathrm{L}'}}(\theta-\delta))
   = \frac{1}{(2\delta)^{1/N}}\begin{bmatrix}
    \mathcal{A}_{[n]}^{\sigma_n^{\mathrm{L}},\sigma_n^{\mathrm{L}'}}(\theta+\delta) & 0 \\
    0 & -\mathcal{A}_{[n]}^{\sigma_n^{\mathrm{L}},\sigma_n^{\mathrm{L}'}}(\theta-\delta)
\end{bmatrix}.
\end{align}
The cyclicity of the trace corresponding to the periodic boundary condition imposed on the light state MPO ensures that the finite difference form in Eq.~\eqref{eq:finitdif} is recovered.
Again, the form of the derivative MPO is not unique, and derivative MPOs with smaller bond dimension may be constructed, e.g., by using compression algorithms~\cite{Schollwock2011}.
However, we use this form for its general purpose applicability even for complicated Hamiltonian parametrisations.
We also note that the bond dimension of this derivative MPO representation is $2D^2$, corresponding to the direct sum structure of each tensor element.

Putting together the forms of the SLD MPO and the derivative MPO from Eqs.~\eqref{eq:sldMPO} and \eqref{eq:derMPO}, we render the variational quantity of interest
\begin{equation}
    F(\theta) = 2\Tr\left[ \dot{\rho}^{\mathrm{L}}(t_\mathrm{fin})  X \right]-\Tr\left[\rho^{\mathrm{L}}(t_\mathrm{fin}) X^2\right]
\end{equation}
as a tensor network quantity, pictorially depicted in Fig.~\ref{fig:scheme}(d).
The evaluation of the QFI then amounts to the simultaneous optimisation over the set $\{X_{[1]},\dots,X_{[N]} \}$ of the quantity of interest $F(\theta)$.

The stringent requirement for simultaneous optimisation of the SLD MPO elements can be relaxed to an element-by-element optimisation, in a variational procedure reminiscent of the density matrix renormalisation group~(DMRG) technique to minimise many-body Hamiltonian ground state energies.
In effect, we first maximise $F(\theta)$ holding all elements of the tensor network diagram, except $X_{[1]}$ fixed, followed by $X_{[2]}$, until $X_{[N]}$, and then loop back to the start in an iterative procedure that is repeated until $F(\theta)$ settles to a constant value (up to a chosen threshold percentage).
The maximisation at each step $n$ is a linear inversion problem, expressed concisely as the optimisation  of the reduced functional
\begin{equation}
    F_n(\theta) = 2\langle b_{[n]} | X_{[n]}\rangle - \langle X_{[n]}|C_{[n]}|X_{[n]}\rangle
\end{equation}
where $\ket{b_{[n]}}$ is the contracted, vectorised complement of the vectorised SLD element $\ket{X_{[n]}}$ in the first term in $F(\theta)$, while $C_{[n]}$ is the matrix complement of vectorised SLD matrices $\ket{X_{[n]}}$, and its Hermitian conjugate. This is depicted graphically in the figure below for $n=2$, adapted from Ref.~\cite{Chabuda2020}.
\begin{center}
	\begin{tikzpicture}[scale=0.5, every node/.style={scale=0.6}]
		\draw[->] (-1,-3.5) -- (-1,-5);
		\draw (-1,-5) node[below] {\LARGE $|b_{[2]}\rangle$};
		\draw (-2.5,-0.5) node {\LARGE $2$};
		\draw (10,-0.5) node {\LARGE $-$};
		\draw (-1,0) to[out=180,in=180,looseness=2] (-1,0.8) -- (8,0.8) to[out=0,in=0,looseness=2] (8,0);
		\draw (-1,-2) to[out=180,in=180,looseness=2] (-1,-1.2) -- (8,-1.2) to[out=0,in=0,looseness=2] (8,-2);

		\draw[line width=1.5mm, white] (2,-5) -- (2,1) to[out=90, in=90,looseness=2] (2+0.75,1) -- (2+0.75,0) to[out=-90, in=90,looseness=2] (2.05,-3) -- (2+0.05,-5) to[out=-90, in=-90,looseness=2] (2+0,-5);
		\draw (2,-5) -- (2,1) to[out=90, in=90,looseness=2] (2+0.75,1) -- (2+0.75,0) to[out=-90, in=90,looseness=2] (2.05,-3) -- (2+0.05,-5) to[out=-90, in=-90,looseness=2] (2+0,-5);

		\foreach \x in {0, 4, 7}
		{
			\draw[line width=1.5mm, white] (\x,-3) -- (\x,1) to[out=90, in=90,looseness=2] (\x+0.75,1) to[out=-90, in=90,looseness=2] (\x+0.75,-3) to[out=-90, in=-90,looseness=2] (\x,-3);
			\draw (\x,-3) -- (\x,1) to[out=90, in=90,looseness=2] (\x+0.75,1) to[out=-90, in=90,looseness=2] (\x+0.75,-3) to[out=-90, in=-90,looseness=2] (\x,-3);
		}
		\draw[line width=1.5mm, white] (-1,0) -- (5,0);
		\draw (-1,0) -- (5,0);
		\draw (5.5,0) node {$\dots$};
		\draw[line width=1.5mm, white] (6,0) -- (8,0);
		\draw (6,0) -- (8,0);

		\draw[line width=1.5mm, white] (-1,-2) -- (1,-2);
		\draw[line width=1.5mm, white] (3,-2) -- (5,-2);
		\draw (-1,-2) -- (1,-2) to[out=0,in=90] (1.95,-4) -- (1.95,-5) -- (2.1,-5) -- (2.1,-4) to[out=90,in=180] (3,-2) -- (5,-2);

		\draw[line width=1.5mm, white] (6,-2) -- (8,-2);
		\draw (6,-2) -- (8,-2);

		\draw (5.5,-2) node {$\dots$};
		
		\foreach \x/\c in {0/{1}, 2/{2}, 4/{3}, 7/{N}}
		{
			\draw[fill=white] (\x-0.6,-0.6) rectangle (\x+0.6,0.6);
			\draw (\x,0) node {$\mathcal{B}_{\theta,[\c]}$};
		}
		\foreach \x/\c in {0/{1}, 4/{3}, 7/{N}}
		{
			\draw[fill=white] (\x-0.6,-2.6) rectangle (\x+0.6,-1.4);
			\draw (\x,-2) node {$X_{[\c]}$};
		}

		\draw[fill=white] (2-0.6,-5-0.6) rectangle (2+0.6,-5+0.6);
		\draw (2,-5) node {$X_{[2]}$};

		\draw[rounded corners, fill=none,dashed] (-1.75,-4) rectangle (8.75,2.0);

		\draw (2,-3.5) node[right] {\LARGE $\alpha$};

		\begin{scope}[shift={(5,-9)}]
		\draw[->] (8,1.5) -- (8,3);
		\draw (8,3) node[above] {\LARGE $C_{[2]}$};
		\draw (-1,0) to[out=180,in=180,looseness=2] (-1,0.8) -- (8,0.8) to[out=0,in=0,looseness=2] (8,0);
		\draw (-1,-2) to[out=180,in=180,looseness=2] (-1,-1.2) -- (8,-1.2) to[out=0,in=0,looseness=2] (8,-2);
		\draw (-1,-4) to[out=180,in=180,looseness=2] (-1,-3.2) -- (8,-3.2) to[out=0,in=0,looseness=2] (8,-4);
		
		\draw[line width=1.5mm, white] (2,-7) -- (2,3) to[out=90, in=90,looseness=2] (2+0.05,3) -- (2.05,1) to[out=-90, in=90,looseness=2] (2+0.75,-1.5) -- (2+0.75,-2.5) to[out=-90,in=90,looseness=2] (2.05,-5) -- (2.05,-6) to[out=-90, in=-90,looseness=2] (2+0.05,-7);

		\draw (2,-7) -- (2,3) to[out=90, in=90,looseness=2] (2+0.05,3) -- (2.05,1) to[out=-90, in=90,looseness=2] (2+0.75,-1.5) -- (2+0.75,-2.5) to[out=-90,in=90,looseness=2] (2.05,-5) -- (2.05,-6) to[out=-90, in=-90,looseness=2] (2+0.05,-7);

		\foreach \x in {0, 4, 7}
		{
			\draw[line width=1.5mm, white] (\x,-5) -- (\x,1) to[out=90, in=90,looseness=2] (\x+0.75,1) to[out=-90, in=90,looseness=2] (\x+0.75,-5) to[out=-90, in=-90,looseness=2] (\x,-5);
			\draw (\x,-5) -- (\x,1) to[out=90, in=90,looseness=2] (\x+0.75,1) to[out=-90, in=90,looseness=2] (\x+0.75,-5) to[out=-90, in=-90,looseness=2] (\x,-5);
		}
		\draw[line width=1.5mm, white] (-1,0) -- (1,0);
		\draw[line width=1.5mm, white] (3,0) -- (5,0);
		\draw (-1,0) -- (1,0) to[out=0,in=-90] (1.95,2) -- (1.95,3) -- (2.1,3) -- (2.1,2) to[out=-90,in=180] (3,0) -- (5,0);

		\draw (5.5,0) node {$\dots$};
		\draw[line width=1.5mm, white] (6,0) -- (8,0);
		\draw (6,0) -- (8,0);

		\draw[line width=1.5mm, white] (-1,-2) -- (5,-2);
		\draw (-1,-2) -- (5,-2);

		\draw[line width=1.5mm, white] (6,-2) -- (8,-2);
		\draw (6,-2) -- (8,-2);

		\draw[line width=1.5mm, white] (-1,-4) -- (1,-4);
		\draw[line width=1.5mm, white] (3,-4) -- (5,-4);

		\draw (-1,-4) -- (1,-4) to[out=0,in=90] (1.95,-5) -- (1.95,-7) -- (2.1,-7) -- (2.1,-5) to[out=90,in=180] (3,-4) -- (5,-4);

		\draw[line width=1.5mm, white] (6,-4) -- (8,-4);
		\draw (6,-4) -- (8,-4);

		\draw (5.5,-2) node {$\dots$};
		\draw (5.5,-4) node {$\dots$};
		\foreach \x/\c in {0/{1}, 4/{3}, 7/{N}}
		{
			\draw[fill=white] (\x-0.6,-0.6) rectangle (\x+0.6,0.6);
			\draw (\x,0) node {$X_{[\c]}$};
			\draw[fill=white] (\x-0.6,-2.6) rectangle (\x+0.6,-1.4);
			\draw (\x,-2) node {$\mathcal{A}_{\theta,[\c]}$};
			\draw[fill=white] (\x-0.6,-4.6) rectangle (\x+0.6,-3.4);
			\draw (\x,-4) node {$X_{[\c]}$};
		}
		\draw[fill=white] (2-0.6,3-0.6) rectangle (2+0.6,3+0.6);
		\draw (2,3) node {$X_{[2]}$};
		\draw[fill=white] (2-0.6,-2-0.6) rectangle (2+0.6,-2+0.6);
		\draw (2,-2) node {$\mathcal{A}_{\theta,[2]}$};
		\draw[fill=white] (2-0.6,-7-0.6) rectangle (2+0.6,-7+0.6);
		\draw (2,-7) node {$X_{[2]}$};
		\draw[rounded corners, fill=none,dashed] (-1.75,-6) rectangle (8.75,2.0);
		\draw (2,-5.5) node[right] {\LARGE $\beta$};
		\draw (2,1.5) node[right] {\LARGE $\alpha$};
		\end{scope}
	\end{tikzpicture}
\end{center}
Setting the first-order derivative with respect to the vector $\ket{X_{[n]}}$ to zero, we then obtain the optimisation condition
\begin{equation}\label{eq:linearAxb}
    \frac{1}{2}(C_{[n]} + C_{[n]}^{T})\,\ket{X_{[n]}} = \ket{b_{[n]}}.
\end{equation}
The matrix $\frac{1}{2}(C_{[n]} + C_{[n]}^{T})$ is often rank-deficient\cite{Chabuda2020}, making the inversion in the above equation delicate.
The rank deficiency is a manifestation of the gauge freedoms in choosing the elements of the tensors $\{X_{[1]},\dots,X_{[N]}\}$.
We control this freedom in the solution by choosing a least-squares solution $\ket{X_{[n]}}$ with the minimum $2$-norm, more details of which will be provided in Appendix~\ref{app:numericaldetails}.

Finally, we note that any least-squares solutions obtained from under-determined linear systems in Eq.~\eqref{eq:linearAxb} need not respect the Hermitian gauge imposed on the initial MPO ansatz for $X$~\cite{Chabuda2020}.
In order to mitigate the effects of this in numerical implementations, we remove any anti-Hermitian component of the solution tensor explicitly by
\begin{equation}
    X_{[n]}^{\sigma_n^{\mathrm{L}},\sigma_n^{\mathrm{L}'}} \rightarrow \frac{1}{2} \left[ X_{[n]}^{\sigma_n^{\mathrm{L}},\sigma_n^{\mathrm{L}'}} + (X_{[n]}^{\sigma_n^{\mathrm{L}'},\sigma_n^{\mathrm{L}}})^*  \right].
\end{equation}
This ensures that the updated SLD MPO is still in Hermitian gauge. 
The threshold condition for terminating the loop through the network, for fixed $D_X$ is set as 
\begin{equation}
    \frac{F_{n+1}(\theta) -F_{n}(\theta)}{F_n(\theta)} < \epsilon_{\mathrm{tol}}
\end{equation}
yielding a converged value $\mathcal{Q}[\rho^{\mathrm{L}}(t_{\mathrm{fin}});D_X]$.
This procedure is then replicated for incremental SLD MPO bond dimensions, and a similar convergennce is demanded, 
\begin{equation}
    \frac{\mathcal{Q}[\rho^{\mathrm{L}}(t_{\mathrm{fin}});D_X+1]-\mathcal{Q}[\rho^{\mathrm{L}}(t_{\mathrm{fin}});D_X]}{\mathcal{Q}[\rho^{\mathrm{L}}(t_{\mathrm{fin}});D_X]} < \epsilon_{\mathrm{tol,bonddim}},
\end{equation}
yielding the final converged value of the MPO-QFI.

\section{Numerical Details of MPO-QFI evaluation}\label{app:numericaldetails}

\begin{table}[ht]
    \centering
\begin{tabular}{lcccccc}
\toprule
 & $\eta$ & $\Delta t$ [$1/\Gamma$] & $\epsilon_{\mathrm{lsq}} $ & $\epsilon_{\mathrm{tol}}$ & $\epsilon_{\mathrm{tol,bonddim}}$ & $R$ \\ 
 \midrule
 & 0.2 & 0.05 & 1E-07 & 1E-02 & 1E-02 & 20 \\
Fig.~\ref{fig:fluorescence}(a) & 0.5 & 0.05 & 1E-07 & 1E-02 & 1E-02 & 20 \\
 & 0.8 & 0.05 & 1E-07 & 1E-02 & 1E-02 & 20 \\
\midrule
 & 0.2 & 0.05 & 1E-07 & 1E-02 & 1E-02 & 51 \\
Fig.~\ref{fig:fluorescence}(b)  & 0.5 & 0.05 & 1E-07 & 1E-02 & 1E-02 & 44 \\
 & 0.8 & 0.05 & 1E-07 & 1E-02 & 1E-02 & 40
\\ 
\bottomrule
\multicolumn{7}{c}{(a)}
\end{tabular}
\hspace{.7cm}
\begin{tabular}{lcccccc}  \toprule
 & $\bar{n}$ & $\Delta t$ [$1/\Gamma$] & $\epsilon_{\mathrm{lsq}} $ & $\epsilon_{\mathrm{tol}}$ & $\epsilon_{\mathrm{tol,bonddim}}$ & $R$ \\ 
 \midrule
 & 1   & 0.0867  & 1E-07 & 1E-02 & 1E-02 & 10 \\
Fig.~\ref{fig:pulsed_Gamma_est}(a)-(b) & 5   & 0.0867  & 1E-07 & 1E-02 & 1E-02 & 10 \\
 & 10  & 0.0867  & 1E-07 & 1E-02 & 1E-02 & 10 \\
 \midrule
 & 1   & 0.1333  & 1E-07 & 1E-02 & 1E-02 & 10 \\
 Fig.~\ref{fig:pulsed_Gamma_est}(c)-(d)  & 5   & 0.1333  & 1E-07 & 1E-02 & 1E-02 & 10 \\
& 10  & 0.1333  & 1E-07 & 1E-02 & 1E-02 & 10 \\ \bottomrule
\multicolumn{7}{c}{(b)}
\end{tabular}
\caption{
    Specific choice of parameters for the numerical examples presented in the main text corresponding to (a) Rabi frequency  $\Omega$  estimation in Sec.~\ref{subsec:rabi_est}, and (b) dipole moment $\Gamma$ estimation in pulsed coherent driving setup in Section \ref{subsec:dipole_est}.}
    \label{tab:numerics}
\end{table}

In Table~\ref{tab:numerics}, we present the numerical parameters used to obtain the MPO-QFIs in Sections~\ref{subsec:rabi_est} and \ref{subsec:dipole_est}.
The critical parameter that controls the accuracy of the evaluated MPO-QFI in both these problems is the size of the coarse-graining parameter $\Delta t$, which determines not just the accuracy of the Trotter expansion (see Eq.~\eqref{eq:Trotterprod}) and the correctness of the Kraus operators (see Eq.~\eqref{eq:Kraus_ops}) that are used to compose the light state MPO $\rho^{\mathrm{L}}(t_{\mathrm{fin}})$, but also controls the success and accuracy of the iterative variational algorithm used to compute the QFI.
More concretely, the size of the $\Delta t$, which relates directly number of sites in the many-body state $N = t_{\mathrm{fin}}/\Delta t$, must be chosen to be sufficiently smaller than the fastest timescales of the observed dynamics, which for the Rabi frequency problem would correspond to the inverse of the Rabi oscillation period $1/\Omega$, whereas for the pulsed spectroscopy problem would be $\mathrm{min}\,(1/\Gamma,T)$.

The inversion step in Eq. (\ref{eq:linearAxb}) was performed using the MATLAB function \verb|lsqminnorm()|, which yields the least square solution of the linear system with the smallest $2$-norm. Empirically, this was found to perform better than the Penrose-Moore pseudoinverse in the context of the variational algorithm, as the magnitude of the entries of the solution SLD matrices $X[n]$ was smaller. The size of $\Delta t$ also affects this inversion. This is because for too small a $\Delta t$, the singular value spectrum of the rank-deficient matrix $\frac{1}{2}(C + C^{T})$ becomes too noisy for a successful inversion. The choice of the rank tolerance parameter $\epsilon_{\mathrm{lsq}}$, below which all singular values are assumed noisy and set to zero, is key in controlling the numerical accuracy of the optimisation.

The contraction of the network is performed using the \verb|ncon()| routine in MATLAB~\cite{pfeifer2014ncon}, while the function \verb|netcon()| is used to find the most efficient order for contractions in the network diagram.
Finally, we repeated the iterative algorithm was replicated $R$ times for more numerical confidence.

\section{Additional numerical results}
\label{app:additional_results}

\begin{figure}
    \includegraphics{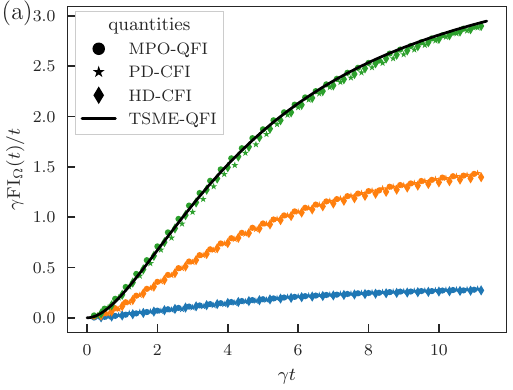}
    \includegraphics{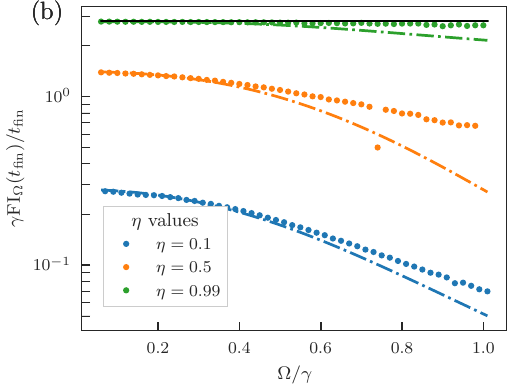}
    \caption{
    QFIs and CFIs for Rabi frequency ($\Omega$) estimation, generically denoted as $\mathrm{FI}_\Omega(t)$ rescaled by the evolution time.
    The markers correspond to different quantities, as in (a); colors correspond to values of the efficiency $\eta$, as in (b).
    Panel (a) shows the dynamics of the various quantities as a function of the evolution time, for a fixed $\Omega = 0.1\gamma$.
    Panel (b)
    shows the the same figures of merit, except the CFIs, evaluated at the time $t_\mathrm{fin}=10/\gamma$, as a function of the Rabi frequency; the dash-dotted lines represents the sub-QFI.
    The MPO-QFI is $\mathcal{Q}[ \rho_\Omega^{\mathrm{L}}]$, the TSME-QFI is $\mathcal{Q}[ \rho_\Omega^{\mathrm{LE}} ]$ which for our problem is also the QFI of $\rho^{\mathrm{LE}}_\Omega $ for $\eta=1$, the PD-CFI is $\mathcal{C}[ p_{\Omega}^{\mathrm{pd}} ]$ and the HD-CFI is $\mathcal{C}[p_\Omega^{\mathrm{hd}}]$.
    The HD-CFI is for a homodyine angle $\varphi = \pi /2$, which is known to be optimal for $\Omega$ estimation~\cite{Kiilerich2016}.
    }
    \label{fig:fluorescence_supp}
\end{figure}

\begin{figure}
    \includegraphics{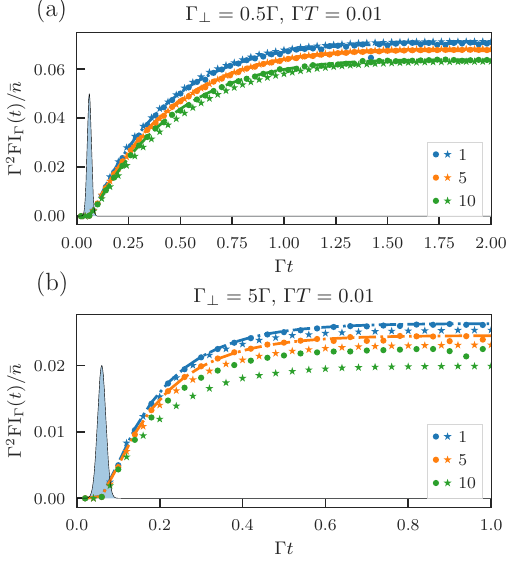}
    \caption{
    QFIs and CFIs for $\Gamma$ estimation with pulsed light, generically denoted as $\mathrm{FI}_\Gamma(t),$ multiplied by $\Gamma^2$ to be adimensional and rescaled by the average number of photons $\bar{n}$ .
  Circles represent the MPO-QFI, stars the PD-CFI, the dash-dotted line the sub-QFI.
  The colors represent different average number of photons $\bar{n}$ for the coherent states.
    The light-blue shapes is a visual aid that represents the pulse profile $|\phi(t)|^2$ (not to scale on the y-axis). The centre of the pulse is at $t_c = 6.5 T$.
    Each panel corresponds to the parameter values shown on top. }
\label{fig:pulsed_Gamma_est_app}
\end{figure}

We present additional numerical results not included in the main text figures, aimed at showcasing new calculations enabled by the MPO framework---specifically, the MPO-QFI and subQFI---rather than offering a comprehensive analysis.
Fig.~\ref{fig:fluorescence_supp} complements Fig.~\ref{fig:fluorescence} by displaying two additional efficiency values, with the $\eta=0.5$ curve repeated for reference. Fig.~\ref{fig:pulsed_Gamma_est_app} parallels Fig.~\ref{fig:pulsed_Gamma_est}, but shows results for a shorter pulse with $\Gamma T = 0.01$.

\end{document}